\documentclass[12pt]{article}

\usepackage[utf8]{inputenc}

\usepackage[ruled, lined, longend]{algorithm2e}
\usepackage{booktabs,multirow}
\usepackage{adjustbox}
\usepackage{tikz}
\usetikzlibrary{quantikz}

\usepackage{wrapfig}
\usepackage{graphicx}
\usepackage[margin=1in]{geometry}

\usepackage{amsmath,amssymb,amsthm}
\newcommand{\Id}{\mathit{Id}}

\newcommand{\CNOT}{\text{CNOT}}
\newcommand{\SWAP}{\text{SWAP}}

\newcommand{\bigO}{\mathcal{O}}
\DeclareMathOperator{\tr}{tr}
\newcommand{\ketbra}[1]{\left|#1\middle\rangle\middle\langle#1\right|}
\newcommand{\dicke}[2]{\ket{\smash{D_{#2}^{#1}}}}
\newcommand{\dickeT}[2]{\bra{\smash{D_{#2}^{#1}}}}
\newcommand{\dickeDM}[2]{\ketbra{\smash{D_{#2}^{#1}}}}
\newcommand{\dickep}[3]{\mathrm{D{#1}{#2}\;\mhyphen\;{#3}}}
\mathchardef\mhyphen="2D

\usepackage[colorlinks=true]{hyperref}

\title{A Divide-and-Conquer Approach\linebreak to Dicke State Preparation%
    \thanks{Research presented in this article was supported by the Laboratory Directed Research and Development program of Los Alamos National Laboratory under project number 20200671DI.%
    \hfill Report LA-UR-21-31138.\newline
    Corresponding authors: \texttt{saktar@nmsu.edu}, \texttt{baertschi@lanl.gov}.}
}
\author{Shamminuj Aktar$^{1,2,*}$,  Andreas Bärtschi$^{2,*}$, \and Abdel-Hameed A. Badawy$^1$, Stephan Eidenbenz$^2$}

\date{
    \small
    $^1$ Klipsch School of Electrical and Computer Engineering, New Mexico State University\\
    $^2$ CCS-3 Information Sciences, Los Alamos National Laboratory
}

\begin{document}
\maketitle

\begin{abstract}
    We present a divide-and-conquer approach to deterministically prepare Dicke states $\dicke{n}{k}$ (i.e., equal-weight superpositions of all $n$-qubit states with Hamming Weight $k$) on quantum computers.
    In an experimental evaluation for up to $n=6$ qubits on IBM Quantum Sydney and Montreal devices, we achieve significantly higher state fidelity compared to previous results~\cite[Mukherjee et.al. TQE'2020]{mukherjee2020preparing},~\cite[Cruz et.al. QuTe'2019]{epfl2019wstate}.
    
    The fidelity gains are achieved through several techniques: 
    Our circuits first ``divide'' the Hamming weight between blocks of $n/2$ qubits, and then ``conquer'' those blocks with improved versions of Dicke state unitaries~\cite[Bärtschi et.al. FCT'2019]{baertschi2019deterministic}. 
    Due to the sparse connectivity on IBM's heavy-hex-architectures, these circuits are implemented for linear nearest neighbor topologies.
    Further gains in (estimating) the state fidelity are due to our use of measurement error mitigation and hardware progress.
\end{abstract}

\section{Introduction}
In quantum computing, Dicke states~\cite{Dicke1954} are a class of highly-entangled quantum states with the rare feature that they are of importance as initial states in quantum algorithms,
in addition to their quantum mechanical property of being highly entangled. The Dicke state $\ket{D^n_k}$ assigns equal non-zero amplitudes of $1/\surd\binom{n}{k}$ to each computational basis state of Hamming weight $k$, where $n$ is the number of qubits and a computational basis state has Hamming weight $k$ if exactly $k$ of the bits take on value $1$,
e.g. $\dicke{4}{2} = \tfrac{1}{\surd{6}}(\ket{1100}+\ket{1010}+\ket{1001}+\ket{0110}+\ket{0101}+\ket{0011}).$
The combinatorial interpretation of a Dicke state is, for example, the set of all feasible solutions of a constraint optimization problem, such as Maximum $k$-Densest Subgraph, which asks for a subset of exactly $k$ vertices of a given input graph with a maximum number of (induced) edges. 

In this paper, we study how well Dicke states can be created on present-day Noisy Intermediate Scale Quantum (NISQ) devices, in particular several IBM Q devices. 
In Section~\ref{sec:circuits}, we propose a novel Divide-and-Conquer approach to designing Dicke state preparation circuits. We present circuits for Dicke states $\dicke{n}{k}$ for $1\leq 2k \leq n\leq 6$  
that are optimized towards minimum circuit depth and {\CNOT}-gate counts. These circuits are our first main result. Compared to earlier work, we achieve reductions in {\CNOT} counts of up to 30 percent; for instance our $\ket{D^4_2}$ circuit requires 10 {\CNOT} gates and has a depth of 11 vs. previously best known values of 12 CNOT gates and a depth of 21~\cite{mukherjee2020preparing}.

We test our circuits on two IBM Q backends (Sydney and Montreal) using several metrics and different compilation options that IBMs QISKIT environment offers.
As our main measure to assess how close (noisy) these NISQ devices actually produce the (ideal pure) quantum state, we calculate the
quantum fidelity of the state~\cite{jozsa1994fidelity} through full state tomography.
This requires the execution of $3^n$ different runs of our circuits which we measure in all possible Pauli bases, followed by a maximum-likelihood estimation of the density matrix representing the prepared mixed state~\cite{james2001tomography}, both of which are natively supported in QISKIT~\cite{qiskit-tomography}. 
Each run needs to be repeated often enough in order to get sufficient statistics on the sampling frequency.

\newcommand{\plog}{\mathrm{poly\,log\,}}
\begin{table}[t]
\scriptsize
    \caption{Different preparation schemes for Dicke and related states. W States are Dicke States of Hamming weight 1, while Symmetric States are superpositions of Dicke States. They can thus be seen as a subset (superset, respectively) of Dicke States. Probabilistic state preparation uses a projective measurement of a $n$-qubit product state into a Hamming weight subspace.
	Quantum compression is more general and is used in reverse for state preparation. 
	$\varepsilon$ denotes the precision of arithmetic circuits.\\
	This work improves circuits and constant factors from the work of Bärtschi and Eidenbenz~\cite{baertschi2019deterministic}. 
	}
    \centering
	\begin{tabular}{@{}llllll@{}}
		\toprule State Prep.
		& Reference					& State Type			& \# {\CNOT} Gates			& Circuit Depth				& \# Ancillas				\\
		\toprule Probabilistic                                                                                          	                                                                               
		& Childs \textit{et.al.} '00~\cite[]{Childs2002}	& Dicke States			& $\bigO(n\ \plog n)$			& $\bigO(n\ \plog n)$			& $\bigO(\log n)$			\\	
		\midrule Deterministic                                                                                          	                                                                               
		& Cruz \textit{et.al.} '18~\cite{epfl2019wstate}	& W States			& $\bigO(n)$				& $\bigO(\log n)$ 			& --					\\
		& Kaye, Mosca '04~\cite{Mosca2001}		& Symmetric States		& $\bigO(n\ \plog(n/\varepsilon))$	& $\bigO(n\ \plog(n/\varepsilon))$	& $\bigO(\log(n/\varepsilon))$		\\			
		& Bärtschi, Eidenbenz '19~\cite{baertschi2019deterministic}& Dicke States	& $\bigO(kn)$				& $\bigO(n)$				& --					\\
		& 						& Symmetric States		& $\bigO(n^2)$				& $\bigO(n)$				& --					\\
		
		\midrule Uncompression                                                                                          	                                                                               
		& Bacon \textit{et.al.} '04~\cite{Bacon2006}	& Schur Transform		& $\bigO(n\ \plog(n/\varepsilon))$	& $\bigO(n\ \plog(n/\varepsilon))$	& $\bigO(\log(n/\varepsilon))$		\\
		& Plesch, Bu\v{z}ek '09~\cite{Plesch2010}	& Symmetric States		& $\bigO(n^2)$				& $\bigO(n^2)$				& --					\\
		
		\bottomrule
	\end{tabular}
	
	\label{tbl:asymptotics}
\end{table}

While we perform the computationally expensive full state tomography in all our experiments, we also explore alternatives. 
We study how well two simpler classical measures that each only require sufficient statistics on a single run upper bound the quantum fidelity. 
These two measures are the measured success probability (informally, how often do we sample a basis state with a Hamming weight $k$)
and the Hellinger fidelity~\cite{hellinger1909}, an analog to quantum fidelity for classical probability distributions.

We test several different QISKIT compiler options that include giving an initial layout of logical to physical qubits, 
using noise adaptive transpilation (IBM's term for compilation~\cite{qiskit-transpiler}), or just a default transpilation, each combined with and without QISKITs measurement error mitigation~\cite{qiskit-mitigation}. 
The details of our experimental setup as well as more formal definitions of our success measures are described in Section~\ref{sec:setup}.

The contributions of this work can be summarized as follows with the details in Section~\ref{sec:results}:
\begin{enumerate}
	\item	Our novel Dicke state circuits lead to the best quantum fidelity results measured to date on IBM machines, e.g., a quantum fidelity for $\ket{D^4_2}$ of $0.87$, which outperforms the previously measured best result of $0.53$~\cite{mukherjee2020preparing}.
	\item	As expected, quantum fidelity for these Dicke states decreases mostly with increasing circuit complexity; 
		it appears to be a largely linear decrease with increasing {\CNOT} counts. One particular notable exception to this rule is the use of noise-adaptive circuit compilation (which sometimes drastically increases {\CNOT} counts).
	\item	Standard measurement error mitigation techniques increase the calculated quantum fidelity by 
		an absolute value of around $0.1$ with no clear dependence on circuit complexity. 
		Relative improvements in achieved fidelity are seen for an increasing number of qubits $n$ and a decreasing Hamming weight $k$ (corresponding to circuits with many measurements but few gates).
	\item	The two IBM Q backends Sydney and Montreal exhibit quite different behaviors, with Montreal generally being much more stable, less susceptible to changes in compilation settings, and better achieved fidelity. 
	\item 	Our two alternative measures of Hellinger fidelity and measured success probability show a similar, albeit flatter linear dependence 
		on {\CNOT} count. Both measures upper bound the quantum fidelity with gaps growing larger with increasing circuit complexity.
\end{enumerate}

Our results show clear technological progress in NISQ devices towards the ability to create entangled states, 
particularly when compared to experiments from one year ago. Parts of these improvements are due to our novel circuit design.

\section{Related work}

Due to their high entanglement, Dicke states~\cite{Dicke1954} have been considered in fields such as quantum game theory~\cite{Oezdemir2007}, 
quantum networking~\cite{Prevedel2009}, quantum metrology~\cite{Toth2012,ouyang2021robust}, quantum error correction~\cite{ouyang2014permutation,ouyang2021permutation} and quantum storage~\cite{ouyang2021quantum}. Their interpretation as superpositions of all feasible states
in Hamming-weight constrained problems have also made them suitable candidates for initial states of adiabatic~\cite{Childs2002} and variational 
combinatorial algorithms~\cite{Hadfield2019,nasa2020XY,cook2020kVC,baertschi2020grover,golden2021grover,golden2022evidence}.
They have been implemented in various platforms such as trapped ions~\cite{Hume2009,Ivanov2013,Lamata2013}, atoms~\cite{Stockton2004,Xiao2007,Shao2010},
photons~\cite{Prevedel2009,Wieczorek2009}, superconducting qubits~\cite{Wu2016}, and others~\cite{johnsson2020geometric,wu2019initializing}.
A Dicke State is defined as an equal superposition of all $n$-bit basis states $x$ of Hamming weight $\mathrm{\rm wt}(x)=k$,
\begin{align}
	\label{eq:dicke}
	\dicke{n}{k} = \binom{n}{k}^{-\frac{1}{2}} \sum\nolimits_{x \in \left\{ 0,1 \right\}^n,\ \mathrm{\rm wt}(x)=k}{\ket{x}}.
\end{align}

Proposed state preparation \emph{circuits} have first relied on arithmetic operations using ancilla registers~\cite{Mosca2001,Bacon2006} until Plesch and Bu\v{z}ek~\cite{Plesch2010} 
gave a quantum compression circuit for symmetric states using $\bigO(n^2)$ gates and depth but no ancillas, see Table~\ref{tbl:asymptotics}. 
This circuit, used in reverse with inverse operations, can be used to prepare Dicke states, because the permutation-invariant symmetric states are
simply superpositions of Dicke states of different Hamming weight $k$.

An improved approach by Bärtschi and Eidenbenz~\cite{baertschi2019deterministic} led to state preparation circuits for symmetric states with $\bigO(n^2)$ gates but linear $\bigO(n)$ depth 
even for Linear Nearest Neighbor (LNN) architectures. 
They also observed that by constricting the symmetric states' Hamming weights to $\leq k$, the number of gates can be decreased to $\bigO(nk) \subset \bigO(n^2)$. 
Mukherjee \textit{et.al.}~\cite{mukherjee2020preparing} later found that narrowing the Hamming weight constriction to exactly $k$, (\textit{i.e.} Dicke States), additional gains 
in lower-order terms of $\bigO(k^2)$ can be made. Unfortunately, both of these gate reduction techniques result in either the need for much higher than LNN connectivity
or the introduction of large constant factors in the gate count~\cite{baertschi2019deterministic}.

In this paper, we circumvent this problem to some extent by dividing a Dicke state preparation into two parts, where we first distribute the Hamming weight $k$ 
over two contiguous blocks of $\lceil n/2 \rceil$ and $\lfloor n/2 \rfloor$ qubits each, before conquering each block with the mentioned LNN scheme. 
\emph{Going forward, we restrict ourselves to Hamming weights $k \leq \lfloor n/2 \rfloor$, as a Dicke state $\dicke{n}{n-k}$ can easily be obtained by flipping all qubits
(applying $X^{\otimes n}$) in the Dicke state $\dicke{n}{k}$}. 
Dividing the Hamming weight in general works best on Ladder architectures, with an increasing {\CNOT} count on LNN connectivities.
For the special case of Dicke states $\dicke{n}{1}$ with Hamming weight 1 (known as W states), this 
method retrieves the best-known {\CNOT} counts of $2n-3$, see Table~\ref{tbl:CNOT-counts}. 

\newcommand{\tph}[1]{\phantom{#1}}
\newcommand{\tss}{\scriptsize}
\begin{table}[t]
    \centering
    \caption{Comparing {\CNOT} counts in our $\dicke{n}{k}$ preparation circuits versus previous approaches 
	    for $k=1$ (Cruz \textit{et.al.} 2018~\cite{epfl2019wstate}) and for $k=2,3$ (Mukherjee \textit{et.al.} 2020~\cite{mukherjee2020preparing}). 
	    Our {\CNOT} counts are attained on LNN architectures, with improved counts for Ladder architectures.
	    $\dicke{n}{n-k}=X^{\otimes n}\dicke{n}{k}$ gives symmetric values for $k>n/2$.
    }
    	\begin{tabular}{@{}lr@{}lrrrrr@{}lrrrrr@{}lrrrr@{}}
		\toprule 
		\		& 		& \multicolumn{5}{c}{\# {\CNOT}s \small{(LNN)}}	& 	& \multicolumn{5}{c}{\# {\CNOT}s \small{(ladder)}} & 	& \multicolumn{5}{c}{prev \# {\CNOT}s}	\\
		\cmidrule(rr){3-7}                                                                                                                                                                  
		\cmidrule(rr){9-13}                                                                                                                                 
		\cmidrule{15-19}                                                                                                                                
				& \ $n=\ $	& 2	& 3	& 4	& 5	& 6	& \ $n=\ $	& 2	& 3	& 4	& 5	& 6	& \ $n=\ $	&  2	& 3	& 4	& 5	& 6	\\
		\midrule                                                                                                                                                                        
		$k=1$		&		& 1	& 3	& 5	& 7	& 9	&		& 1	& 3	& 5	& 7	& 9	&		& 1	& 3	& 5	& 7	& 9	\\
		$k=2$		&		& 	& 	& 10	& 17	& 24	&		& 	& 	& 7	& 14	& 21	&		& 	& 	& 12	& 20	& 28	\\
		$k=3$		&		& 	& 	& 	& 	& 32	&		& 	& 	& 	& 	& 23	&		& 	& 	& 	& 	& 33	\\
		\bottomrule
	\end{tabular}
	\label{tbl:CNOT-counts}
\end{table}

Previous works~\cite{zhang2021low,araujo2021divide} have also included divide-and-conquer strategies for preparing arbitrary quantum states with reduced circuit depth.
Recently, preparation circuits have been proposed for sparse quantum states~\cite{de2021double, malvetti2021quantum, zhang2022quantum} in terms of the number of non-zero state vector entries, requiring logarithmic depth but a quasi-linear number of ancillary qubits and gates. Dicke states can be treated as sparse quantum states for constant Hamming weight, but the number of non-zero entries in Dicke states scales exponentially as $\binom{n}{k}$ with increasing Hamming weight $k$. Hence, sparse state preparation circuits quickly become infeasible for Dicke states preparation.

\begin{figure}[t!]
	\centering
	\includegraphics[width=\linewidth]{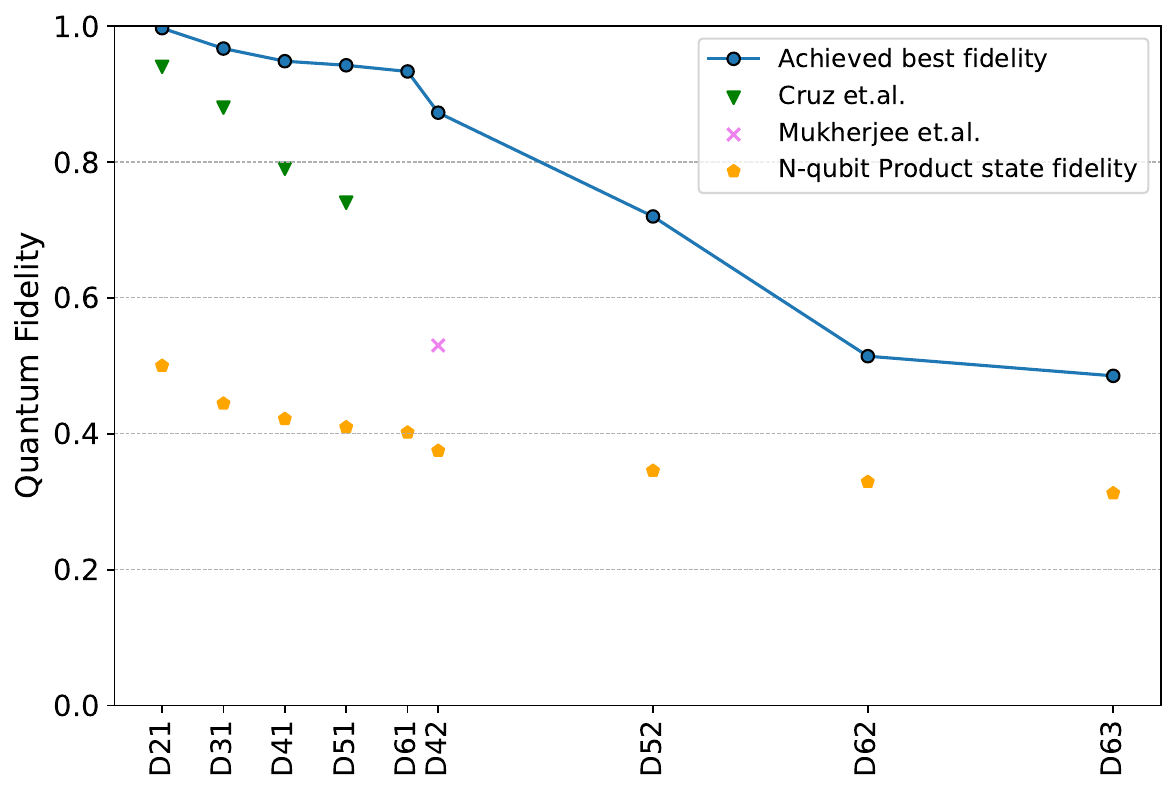}
        \caption{Comparison of our highest measured quantum fidelities with three benchmarks: 
		measured fidelities for five \emph{W states} $\dicke{n}{1}$ (Cruz \textit{et.al.} 2018~\cite{epfl2019wstate})
		and the \emph{Dicke state} $\dicke{4}{2}$ (Mukherjee \textit{et.al.} 2020~\cite{mukherjee2020preparing}), 
		as well as the best possible fidelity among all $n$-qubit \emph{product states}.
		Dicke states are distributed horizontally according to our {\CNOT} counts.
	}
        \label{fig:fidelity-comparison}
\end{figure}

Additionally, we note a probabilistic state preparation approach~\cite{Childs2002} that yields Dicke states with success probability $\binom{n}{k}(\tfrac{k}{n})^k(1-\tfrac{k}{n})^{n-k}$ by preparing the 
symmetric $n$-qubit product state $(\smash{\sqrt{1-k/n}\ket{0} + \sqrt{k/n}\ket{1}})^{\otimes n}$, followed by the addition~\cite{Chuang2000} of the Hamming weight into an ancilla register with $\log n$ qubits and 
a projective measurement thereof. We use the first part of this idea by using the fidelity between such a product state and the Dicke state $\dicke{n}{k}$ (corresponding to the success probability of the projective measurement) as an additional benchmark for our quantum fidelity results, see Figure~\ref{fig:fidelity-comparison}. We additionally prove that no other pure $n$-qubit product state 
has higher fidelity / squared state overlap with the corresponding Dicke state.

\section{Divide-and-Conquer Circuits}
\label{sec:circuits}

\newcommand\mycommfont[1]{\footnotesize\ttfamily{#1}}
\SetCommentSty{mycommfont}
\begin{algorithm}[t]
    \caption{Preparing Dicke State $\dicke{n}{k}$}
        
         \textbf{Step-1: Divide} the Hamming weight $k$\newline between two registers with $\lfloor n/2\rfloor$ and $\lceil n/2 \rceil$ qubits.
         
        \tcc{On a Ladder architecture, this requires $\bigO(k)$ circuit depth \& gate count.\newline
		On a LNN architecture, this requires $\bigO(k)$ circuit depth with $\bigO(k^2)$ gates.
	}
        \tcc{We optimized the 'Divide' circuits specifically for LNN to account for the sparse connectivity of IBM devices, 
		and describe the tuning of the circuits in Figures \ref{fig:divide-conquer-ladder}$\mhyphen$\ref{fig:divide-LNN-D63-32}.
	}
        \textbf{Step-2: Conquer} the blocks\newline by implementing Dicke state unitaries on both registers.
        
        \tcc{Our implementation requires $\bigO(n)$ circuit depth and $\bigO(kn)$ gate count.}
        \label{algorithm-1}
\end{algorithm}

In this section, we give an overview of the techniques behind our Dicke state preparation circuits from Algorithm~\ref{algorithm-1}. They can best be described as a Divide-and-Conquer approach, where the ``Conquer'' part is based on improved versions of \emph{Dicke state unitaries}~\cite{baertschi2019deterministic} and the initial ``Divide'' part uses the new idea of distributing the Hamming weight across two contiguous blocks of qubits. We explain the implementation of Dicke state unitaries, distinguishing between big-endian notation (most significant bit -- the ``big end'' -- on the left of the bit string) and little-endian notation (reversed ordering) for their input encodings. We first briefly review Dicke state unitaries $U_{n,k}$ which denote any unitary satisfying
\begin{align}
	\label{eq:unitary}
	\forall\, k' \leq k\colon U_{n,k}\ket{0^{n-k'}1^{k'}} = \dicke{n}{k'}.   
\end{align}
That is, $U_{n,k}$ takes as possible input a zero-padded unary encoding of any Hamming weight $k' \leq k$ to prepare the Dicke state $\dicke{n}{k'}$.

Previous approaches to prepare $\dicke{n}{k}$ simply fed the input $\ket{0^{n-k}1^{k}}$ to  $U_{n,k}$~\cite{epfl2019wstate,baertschi2019deterministic}. 
In this paper, we present a new approach in which we first divide the Hamming weight into all possible combinations of $k_1 + k_2 = k$ across two blocks of $n_1 + n_2 = n$ qubits, followed by two parallel Dicke state unitaries $U_{n_1,k},\ U_{n_2,k}$ on these blocks. By doing so, we must give the correct weights to the corresponding unary encodings $\ket{0^{n_1-k_1}1^{k_1}}\ket{0^{n_2-k_2}1^{k_2}}$. 
Since there are $\binom{n_1}{k_1}\binom{n_2}{k_2}$ computational basis states with Hamming weight $k_1$ in the $n_1$-bit-prefix and Hamming weight $k_2$ in the $n_2$-bit-suffix, the ``Divide'' part consists of preparing the state
\begin{align}
	\label{eq:divide}
	\frac{1}{\surd{\binom{n}{k}}} \sum_{k_1+k_2=k} \sqrt{\tbinom{n_1}{k_1}\tbinom{n_2}{k_2}} \ket{0^{n_1-k_1}1^{k_1}}\ket{0^{n_2-k_2}1^{k_2}}.
\end{align}

We next discuss Dicke state unitaries -- the ``Conquer'' part of our circuits -- and our improved implementations thereof in detail, before giving the ``Divide'' part for both Ladder and LNN architectures.

\subsection{``Conquer'': Dicke State Unitaries}
\label{conquer-unitaries}

\newcommand{\ryangle}[3]{\underset{\scriptstyle\smash{\surd#1}}{#3\theta#2}}
\newcommand{\rygate}[3]{\gate{\ryangle{#1}{#2}{#3}}}
\begin{figure*}[t!]
	\centering
	\begin{adjustbox}{width=\linewidth}
		\begin{quantikz}[row sep={24pt,between origins},execute at end picture={
					\node[fit=(\tikzcdmatrixname-1-18)(\tikzcdmatrixname-2-22),draw,dashed,thick,rounded corners,inner xsep=10pt,inner ysep=12pt,xshift=-6pt,yshift=4pt] {};
					\node[label={[xshift=-4pt,yshift=-5pt]$U_{2,2}^{be}$}] at (\tikzcdmatrixname-1-18) {};%
				}]
			\lstick[3]{\rotatebox{90}{$\ket{0^{3-k'}1^{k'}}$}}
			q_1	& \qw		& \targ{}	& \rygate{2/3}{/2}{}	& \targ{}	& \qw		& \qw	q_2	& \qw		& \qw			& \ctrl{1}	& \qw			& \qw		& \qw			& \ctrl{1}	& \qw			& \qw		& \qw q_2	& \qw		& \targ{}	& \rygate{1/2}{/2}{}	& \targ{}	& \qw		& \qw	q_3\rstick[3]{\rotatebox{90}{$\dicke{3}{k'}$}}	\\
			q_2	& \gate{\pi/2}	& \ctrl{-1}	& \rygate{2/3}{/2}{}	& \ctrl{-1}	& \gate{-\pi/2}	& \qw	q_1	& \ctrl{1}	& \rygate{2/3}{/4}{-}	& \targ{}	& \rygate{2/3}{/4}{}	& \targ{}	& \rygate{2/3}{/4}{-}	& \targ{}	& \rygate{2/3}{/4}{}	& \ctrl{1}	& \qw q_3	& \gate{\pi/2}	& \ctrl{-1}	& \rygate{1/2}{/2}{}	& \ctrl{-1}	& \gate{-\pi/2}	& \qw	q_2						\\
			q_3	& \qw		& \qw		& \qw			& \qw		& \qw		& \qw	q_3	& \targ{}	& \qw			& \qw		& \qw			& \ctrl{-1}	& \qw			& \qw		& \qw			& \targ{}	& \qw q_1	& \qw		& \qw		& \qw			& \qw		& \qw		& \qw	q_1
		\end{quantikz}
	\end{adjustbox}
	\caption{Implementation of the big-endian Dicke State Unitary $U_{3,3}^{be}=U_{3,2}^{be}$ which takes as input $\ket{0^{3-k'}1^{k'}}$ for any $k'$
	    to prepare $\dicke{3}{k'}$. A small-endian Dicke State Unitary $U_{3,3}^{le}$ has the wire order reversed and accepts inputs $\ket{1^{k'}0^{3-k'}}$ instead.\newline
		The circuit incorporates {\SWAP}s into the actions on the three logical qubits $q_3q_2q_1$: $\ket{001} \mapsto \surd\tfrac{1}{3}\ket{001} + \surd\tfrac{2}{3}\ket{010}$
		and $\ket{011} \mapsto \surd\tfrac{2}{3}\ket{011} + \surd\tfrac{1}{3}\ket{110}$, as well as
		$\ket{01.} \mapsto \surd\tfrac{1}{2}\ket{01.} + \surd\tfrac{1}{2}\ket{10.}$ in the unitary $U_{2,2}^{be}$.
		This changes the ordering of the logical qubits throughout the circuit as shown, where the final order does not matter due to the symmetry of 
		the final state $\dicke{3}{k}$.%
	}
	\label{fig:U33}
	\vspace{-2ex}
\end{figure*}
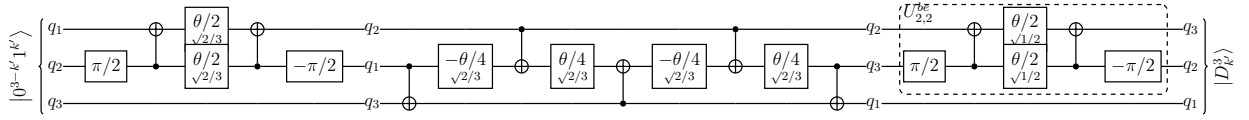

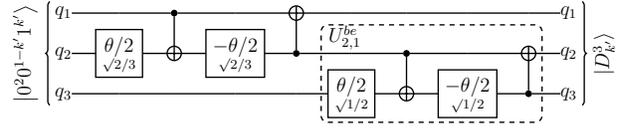
\begin{wrapfigure}{r}{0.5\linewidth}
	\centering
	\vspace{2ex}
	\begin{adjustbox}{width=\linewidth}
		\begin{quantikz}[row sep={24pt,between origins},execute at end picture={
					\node[fit=(\tikzcdmatrixname-2-6)(\tikzcdmatrixname-3-9),draw,dashed,thick,rounded corners,inner xsep=12pt,inner ysep=16pt,xshift=-6pt,yshift=1pt] {};
					\node[label={[xshift=-4pt,yshift=-6pt]$U_{2,1}^{be}$}] at (\tikzcdmatrixname-2-6) {};%
				}]
			\lstick[3]{\rotatebox{90}{$\ket{0^20^{1-k'}1^{k'}}$}}
			q_1	& \qw			& \ctrl{1}	& \qw			& \targ{}   & \qw			& \qw		& \qw			& \qw		& \qw	q_1\rstick[3]{\rotatebox{90}{$\dicke{3}{k'}$}}	\\
			q_2	& \rygate{2/3}{/2}{}	& \targ{}	& \rygate{2/3}{/2}{-}	& \ctrl{-1}	& \qw			& \ctrl{1}	& \qw			& \targ{}	& \qw	q_2						\\
			q_3	& \qw			& \qw		& \qw			& \qw		& \rygate{1/2}{/2}{}	& \targ{}	& \rygate{1/2}{/2}{-}	& \ctrl{-1}	& \qw	q_3
		\end{quantikz}
	\end{adjustbox}
	\caption{Implementation of the Dicke State Unitary $U_{3,1}^{be}$, which takes as input the padded big-endian unary encoding of $k'\leq 1$, 
		(i.e., $\ket{0^2 0^{1-k'}1^{k'}}$) to create the Dicke State $\dicke{3}{k'}$.
	    Labeled gates are $R_y$-rotations, where $\theta_{\surd x/y} = 2\cos^{-1}\surd\tfrac{x}{y}$.
	    The first half of the circuit implements $\ket{001} \mapsto \surd \tfrac{1}{3}\ket{001} + \surd\tfrac{2}{3}\ket{010}$, 
	    the second half the recursively used unitary $U_{2,1}^{be}$.
	}
	\vspace{-2ex}	
	\label{fig:U31}
\end{wrapfigure}

Dicke state unitaries $U_{n,k}$~\cite[\eqref{eq:unitary}]{baertschi2019deterministic} implement the induction
\[	\dicke{n'}{k'} = \sqrt{\tfrac{k'}{n'}} \dicke{n'-1}{k'-1} \otimes\ket{1} + \sqrt{\tfrac{n'-k'}{n'}} \dicke{n'-1}{k'} \otimes\ket{0} \]
for all $n' \leq n,\ k' \leq \min\{k,n'\}$.
Given that $U_{n,k}$ gets as input $\ket{0^{n-k'}1^{k'}}$, it is enough to implement for all $k' \leq k$ the gate
$\ket{0^{n-k'}1^{k'}} \mapsto \surd{\tfrac{k'}{n}} \ket{0^{n-k'}1^{k'}} + \surd{\tfrac{n-k'}{n}} \ket{0^{n-k'-1}1^{k'}0}$ 
followed by a recursive call to the smaller unitary $U_{n-1,k}$, see Figures~\ref{fig:U33},~\ref{fig:U31}. 

For $U_{n,1}$ and $k'=0,1$, this involves {\CNOT}s and a controlled $R_y(2\cos^{-1}\surd\tfrac{1}{n})$-rotation, which we compile down to two single-qubit $R_y(\pm\tfrac{1}{2} \cdot 2\cos^{-1}\surd\tfrac{n-1}{n})$ rotations and two {\CNOT}s. This places $U_{n-1,1}$ on the lower $n-1$ wires, see Figure~\ref{fig:U31}, and thus can be implemented even on LNN architectures.

For $U_{n,k}$ with $k'>1$, we have longer blocks of excitations in the state $\ket{0^{n-k'}1^{k'-1}1^1}$ which has to be mapped~to 
$\surd{\tfrac{k'}{n}} \ket{0^{n-k'}1^{k'-1}1^1} + \surd{\tfrac{n-k'}{n}} \ket{0^{n-k'-1}1^11^{k'-1}0}$.
Still, only three qubits are necessary for this gate, the two qubits changing 0/1-values and the leading 1 as a control. 
The gate can be compiled to {\CNOT}s and a doubly-controlled $R_y(2\cos^{-1}\surd\tfrac{k'}{n})$-rotation~\cite{baertschi2019deterministic}. 
In order to implement the gate on three contiguous blocks of qubits, we incorporate {\SWAP}s which move the lowest-endian qubit from top to bottom, 
see Figure~\ref{fig:U33}. This keeps the {\CNOT} count for gates with $k'=1$ intact, while for $k'>1$ we get an implementation
with 5 {\CNOT}s and 4 $R_y(\pm \tfrac{1}{4}\cdot 2 \cos^{-1}\surd\tfrac{k'}{n})$-rotations.
Finally, we recurse from $U_{n,k}$ to $U_{n-1,k}$ (where in case $k=n$, we use $U_{n,n} = U_{n,n-1}$ to get $U_{n-1,n-1}$).
The detailed shifting-down procedure for the lowest-endian qubit places $U_{n-1,k}$ on the upper $n-1$ wires. 

Finally, we note that our descriptions assumed big-endian inputs $\ket{0^{n'-k'}1^{k'}}$ and thus we further specify 
the given Dicke state unitaries by $U_{n,k}^{be}$. A little-endian version $U_{n,k}^{le}$ accepting little-endian inputs $\ket{1^{k'}0^{n'-k'}}$
is achieved by simply reversing the wire order. The ``Divide'' procedure discussed next will result in an application of both types of Dicke state unitaries, namely $U_{\lfloor n/2 \rfloor,k}^{be}$ and $U_{\lceil n/2 \rceil,k}^{le}$.

\subsection{``Divide'' for Ladder Topologies}

In hardware with ladder architectures, we have two rows of qubits which are provided with 2-qubit gates towards their horizontal nearest neighbors, 
as well as with their direct neighbor below, respectively above. Such topologies can be found, for example, as subgraphs of Google's sycamore processor
and until recently, among IBM Q's devices.
The ``Divide'' technique in the following can also be applied (with constant overhead) to sparser connectivities such as IBM Q's heavy-hex lattices or 
serve as an inspiration for an adaption to LNN architectures discussed in the next subsection.

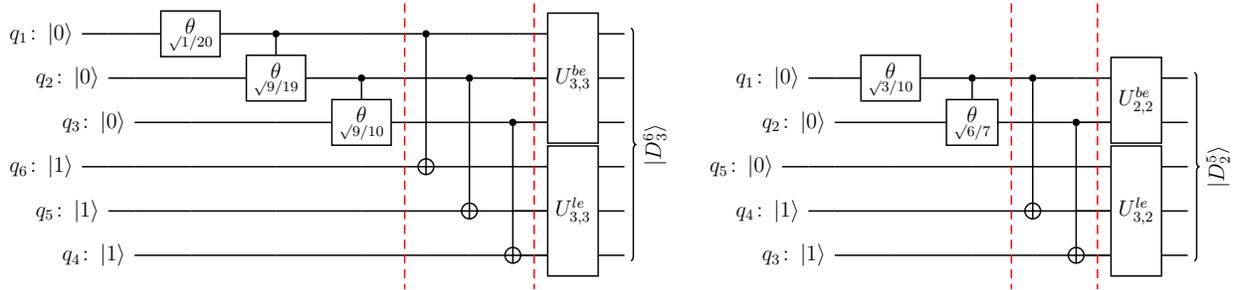
\begin{figure*}[t!]
	\centering
	\begin{adjustbox}{width=\linewidth}
		\begin{quantikz}[row sep={24pt,between origins},execute at end picture={}]
			\lstick{$q_1\colon\ket{0}$}	& \qw				& \qw				& \rygate{1/20}{}{}	& \ctrl{1}		& \qw\slice{}		& \ctrl{3}	& \qw		& \qw\slice{}	& \gate[3]{U_{3,3}^{be}}& \qw\rstick[6]{\rotatebox{90}{$\dicke{6}{3}$}} & 	& 	& 	& 	& 	& 				&				& 				& 			& \slice{}		& 		& \slice{}	& 			& 						\\ 
							& \lstick{$q_2\colon\ket{0}$}	& \qw				& \qw			& \rygate{9/19}{}{}	& \ctrl{1}		& \qw		& \ctrl{3}	& \qw		& \qw			& \qw					 	& 	& 	& 	& 	& 	& 				& \lstick{$q_1\colon\ket{0}$}	& \qw				& \rygate{3/10}{}{}	& \ctrl{1}		& \ctrl{3}	& \qw		& \gate[2]{U_{2,2}^{be}}& \qw\rstick[5]{\rotatebox{90}{$\dicke{5}{2}$}}	\\
							& 				& \lstick{$q_3\colon\ket{0}$}	& \qw			& \qw			& \rygate{9/10}{}{}	& \qw		& \qw		& \ctrl{3}	& \qw			& \qw 						& 	& 	& 	& 	& 	& 				&				& \lstick{$q_2\colon\ket{0}$}	& \qw			& \rygate{6/7}{}{}	& \qw		& \ctrl{3}	& \qw			& \qw					 	\\
			\lstick{$q_6\colon\ket{1}$}	& \qw				& \qw				& \qw			& \qw			& \qw			& \targ{}	& \qw		& \qw		& \gate[3]{U_{3,3}^{le}}& \qw 						& 	& 	& 	& 	& 	& \lstick{$q_5\colon\ket{0}$}	& \qw				& \qw				& \qw			& \qw			& \qw		& \qw		& \gate[3]{U_{3,2}^{le}}& \qw 						\\
							& \lstick{$q_5\colon\ket{1}$}	& \qw				& \qw			& \qw			& \qw			& \qw		& \targ{}	& \qw		& \qw			& \qw 						& 	& 	& 	& 	& 	& 				& \lstick{$q_4\colon\ket{1}$}	& \qw				& \qw			& \qw			& \targ{}	& \qw		& \qw			& \qw 						\\
							&				& \lstick{$q_4\colon\ket{1}$}	& \qw			& \qw			& \qw			& \qw		& \qw		& \targ{}	& \qw			& \qw 						& 	& 	& 	& 	& 	& 				& 				& \lstick{$q_3\colon\ket{1}$}	& \qw			& \qw			& \qw		& \targ{}	& \qw			& \qw 						
		\end{quantikz}
	\end{adjustbox}
	\caption{%
		Preparations of Dicke States on ladder topologies with a divide-and-conquer approach: The Hamming weight $k$ is divided between 
		an upper big-endian row of $n_1 = \lfloor n/2 \rfloor$ qubits and a lower, little-endian row of $n_2 = \lceil n/2 \rceil$ qubits. 
		The resulting superposed entangled row input weights are conquered with the unitaries $U_{n_1,k}^{be}$ and $U_{n_2,k}^{le}$, respectively. \newline
		(LEFT) First we construct the weighted big-endian superposition $\surd\tfrac{1}{20}\left(\surd1\ket{000} + \surd9\ket{001} + \surd9\ket{011} +\surd1\ket{111}\right)$ 
		in the upper row. Note that the numerators $1,9,9$ and the suffix sums $20,19,10$ appear as terms in the angle arguments. Next we subtract these weights 
		from the little-endian Hamming weight $k=3$ in the lower row. Finally we use Dicke State unitaries on both rows to construct $\dicke{6}{3}$. 
		(RIGHT) Same approach for $\dicke{5}{2}$, where we first construct $\surd\tfrac{1}{10}\left(\surd{3}\ket{00}+\surd{6}\ket{01}+\surd{1}\ket{11}\right)$ in the 
		upper row and subtract from Hamming weight $k=2$ in the lower row.%
	}
	\label{fig:divide-conquer-ladder}
\end{figure*}

Recall that we restrict ourselves to Hamming weight values $k\leq n/2$ due to the symmetry $\dicke{n}{k} = X^{\otimes}\dicke{n}{k}$. 
We divide our $n$ qubits into $n_1 = \lfloor n/2 \rfloor$ qubits $q_1, \ldots, q_{n_1}$ on the upper row and 
$n_2 = \lceil n/2 \rceil$ qubits $q_n, \ldots, q_{n_1+1}$ (in this order) on the lower row, such that qubit $q_i,\ i \leq n_1$ 
is vertically connected to qubit $q_{2n_1-i+1}$, see Figure~\ref{fig:divide-conquer-ladder}.

To get the desired state from Equation~\eqref{eq:divide}, we first prepare 
\[ \frac{1}{\surd{\binom{n}{k}}} \sum_{k_1+k_2=k} \sqrt{\tbinom{n_1}{k_1}\tbinom{n_2}{k_2}} \ket{0^{n_1-k_1}1^{k_1}}\ket{0^{n_2-k}1^{k}}. \]

Let $x_i = \binom{n_1}{i}\binom{n_2}{k-i}$ denote the number of bitstrings with Hamming weight $i$ among the first half of $n_1$ digits 
and Hamming weight $k-i$ among the second half of $n_2 = n-n_1$ digits. 
Of the sequence $x_0,\ldots,x_k$, let $s_i = x_i + \ldots + x_k$ denote the suffix sums of the $x_i$-values.
Then, we can construct the superposition $\sum_{k_1+k_2=k} \sqrt{\tbinom{n_1}{k_1}\tbinom{n_2}{k_2}} \ket{0^{n_1-k_1}}$
(modulo proper normalization) one term at a time with controlled $R_y(2\cos^{-1}\surd \tfrac{x_i}{s_i})$-rotations \emph{up until $i=k-1$};
consecutively rotating qubits one-by-one controlled on the value of previous, smaller-endian qubits, see Figure~\ref{fig:divide-conquer-ladder}.
The compilation of a $\mathit{CR}_y$-gate into single-qubit and {\CNOT} gates usually takes 2 {\CNOT}s, however for a classical 0-or-1 target, 
it can be done with 1 {\CNOT} only (with compilation based on knowledge of the classical value).

We can then subtract (in superposition) the Hamming weight $k_1$ encoded in big-endian unary in the first half of $n_1$ qubits 
from the $\ket{0^{n_2-k}1^{k}}$ unary encoding of $k$ in the second half of $n_2$ qubits. This can be done with 1 {\CNOT} for each pair 
$(q_i, q_{2n_1-i+1})$.
Taking also into account the number of {\CNOT}s of a big-endian unitary $U_{n_1,k}^{be}$ on the upper $n_1$ qubits and 
a little-endian unitary $U_{n_2,k}^{be}$ on the lower, reversely ordered $n_2$ qubits, we get the {\CNOT} counts for Ladder architectures 
in Table~\ref{tbl:CNOT-counts}.

\renewcommand{\ryangle}[3]{\underset{\scriptscriptstyle\smash{\surd#1}}{#3\theta#2}}
\renewcommand{\rygate}[3]{\gate[style={inner sep=0pt}]{\scriptstyle\ryangle{#1}{#2}{#3}}}
\begin{figure*}[t!]
	\begin{adjustbox}{width=0.8\linewidth}
		\begin{quantikz}[row sep={24pt,between origins},execute at end picture={
				\node[fit=(\tikzcdmatrixname-2-10)(\tikzcdmatrixname-3-13),draw,dashed,thick,rounded corners,inner sep=2pt,label={[yshift=-4pt]above:$\mathit{CR}_y(\theta_{\surd 9/10})+{\SWAP}$}] {};
				\node[fit=(\tikzcdmatrixname-5-7)(\tikzcdmatrixname-6-8),draw,dashed,thick,rounded corners,inner sep=4pt,label={[yshift=0pt]above:{\SWAP}}] {};
			}]
			\lstick{$q_3\colon\ket{0}$}	& \qw			& \qw		& \qw		& \qw		& \qw			& \targ{}	& \ctrl{1}	& \qw q_1	& \qw			& \qw		& \qw		& \qw			& \qw   	& \qw q_1			& \gate[3]{U_{3,3}^{be}}	& \qw\rstick[6]{\rotatebox{90}{$\dicke{6}{3}$}}	\\	 
			\lstick{$q_2\colon\ket{0}$}	& \rygate{10/19}{/2}{}	& \qw		& \ctrl{1}	& \targ{}	& \qw			& \ctrl{-1}	& \targ{}	& \qw q_3	& \rygate{1/10}{/2}{}	& \ctrl{1}	& \targ{}	& \qw			& \qw   	& \qw q_2			& \qw				& \qw						\\	 
			\lstick{$q_1\colon\ket{0}$}	& \rygate{1/20}{}{}	& \ctrl{1}	& \targ{}	& \ctrl{-1}	& \rygate{10/19}{/2}{-}	& \ctrl{1}	& \qw   & \qw q_2	& \qw			& \targ{}	& \ctrl{-1}	& \rygate{1/10}{/2}{-}	& \ctrl{1}	& \qw q_3		& \qw				& \qw						\\	 
			\lstick{$q_6\colon\ket{0}$}	& \qw			& \targ{}	& \ctrl{1}	& \targ{}	& \qw			& \targ{}   & \qw	& \qw q_5	& \targ{}		& \ctrl{1}	& \targ{}	& \qw			& \targ{}   & \qw q_4		& \gate[3]{U_{3,3}^{be}}	& \qw						\\	 
			\lstick{$q_5\colon\ket{0}$}	& \targ{}		& \qw		& \targ{}	& \ctrl{-1}	& \targ{}		& \ctrl{1}	& \targ{}	& \qw q_4	& \qw			& \targ{}	& \ctrl{-1}	& \qw			& \qw       & \qw q_5		& \qw				& \qw						\\	 
			\lstick{$q_4\colon\ket{0}$}	& \targ{}	 	& \qw		& \qw		& \qw		& \qw			& \targ{}	& \ctrl{-1}	& \qw q_6	& \qw			& \qw		& \qw		& \qw			& \qw       & \qw q_6	    & \qw				& \qw			 	
		\end{quantikz}
	\end{adjustbox}
	\caption{%
		Implementation of Dicke state $\dicke{6}{3}$ on LNN starting from a reverse ordering of the qubit blocks $(q_1,q_2,q_3)$ and $(q_4,q_5,q_6)$:
		We start with neighboring qubits $q_1,q_6$, implement their interaction and then swap them to the far ends of the circuit. 
		While doing so, we use the fact that controlled-$R_y$s and {\SWAP}s target qubits in a classical state, which reduces implementations of {\SWAP} and 
		$\mathit{CR}_y+{\SWAP}$ interactions to two {\CNOT}s each.\newline
		For the ``Divide'' part we get a {\CNOT} count of 15 and a {\CNOT} depth of 8, while for the ``Conquer'' part in the big-endian unitaries
		$U_{3,3}^{be}$ we have 9 {\CNOT}s. Overall we get a {\CNOT} count of 33 and a {\CNOT} depth of 17.
	}
	\label{fig:divide-LNN-D63-33}
\end{figure*}
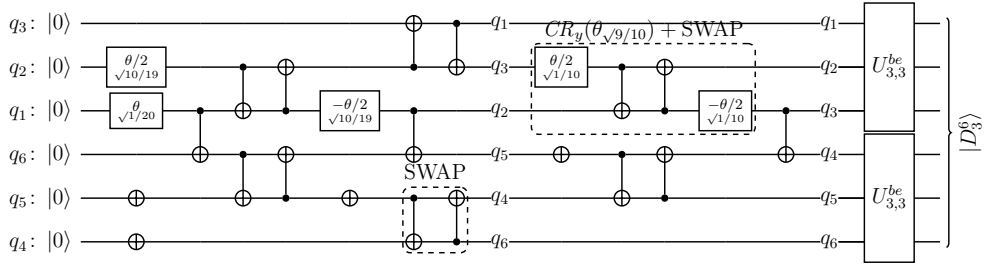

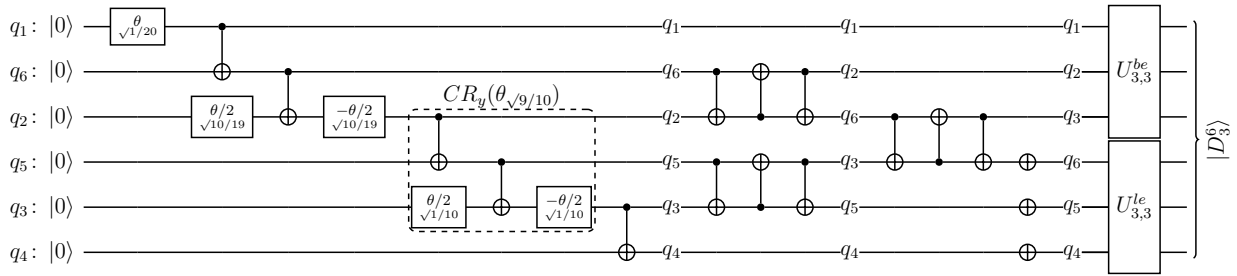
\begin{figure*}[t!]
	\begin{adjustbox}{width=\linewidth}
		\begin{quantikz}[row sep={24pt,between origins},execute at end picture={
				\node[fit=(\tikzcdmatrixname-3-6)(\tikzcdmatrixname-5-8),draw,dashed,thick,rounded corners,inner xsep=8pt,xshift=-6pt,inner ysep=2pt,label={[yshift=-4pt]above:$CR_y(\theta_{\surd 9/10})$}] {};
			}]
			\lstick{$q_1\colon\ket{0}$}	& \rygate{1/20}{}{}	& \ctrl{1}		& \qw		& \qw			& \qw			& \qw		& \qw 			& \qw       	& \qw q_1	& \qw		& \qw		& \qw		& \qw q_1	& \qw		& \qw		& \qw		& \qw		& \qw q_1	& \gate[3]{U_{3,3}^{be}}	& \qw\rstick[6]{\rotatebox{90}{$\dicke{6}{3}$}}	\\	 
			\lstick{$q_6\colon\ket{0}$}	& \qw			& \targ{}		& \ctrl{1}	& \qw			& \qw			& \qw		& \qw 			& \qw      	& \qw q_6	& \ctrl{1}	& \targ{}	& \ctrl{1}	& \qw q_2	& \qw		& \qw		& \qw		& \qw		& \qw q_2	& \qw				& \qw						\\	 
			\lstick{$q_2\colon\ket{0}$}	& \qw			& \rygate{10/19}{/2}{}	& \targ{}	& \rygate{10/19}{/2}{-}	& \ctrl{1}		& \qw		& \qw 			& \qw       	& \qw q_2	& \targ{}	& \ctrl{-1}	& \targ{}	& \qw q_6	& \ctrl{1}	& \targ{}	& \ctrl{1}	& \qw		& \qw q_3	& \qw				& \qw						\\	 
			\lstick{$q_5\colon\ket{0}$}	& \qw			& \qw			& \qw		& \qw			& \targ{}		& \ctrl{1}	& \qw 			& \qw       	& \qw q_5	& \ctrl{1}	& \targ{}	& \ctrl{1}	& \qw q_3	& \targ{}	& \ctrl{-1}	& \targ{}	& \targ{}	& \qw q_6	& \gate[3]{U_{3,3}^{le}}	& \qw						\\	 
			\lstick{$q_3\colon\ket{0}$}	& \qw			& \qw			& \qw		& \qw			& \rygate{1/10}{/2}{}	& \targ{}	& \rygate{1/10}{/2}{-}	& \ctrl{1}	& \qw q_3 	& \targ{}	& \ctrl{-1}	& \targ{}	& \qw q_5	& \qw		& \qw		& \qw		& \targ{}	& \qw q_5	& \qw				& \qw						\\	 
			\lstick{$q_4\colon\ket{0}$}	& \qw		 	& \qw			& \qw		& \qw			& \qw			& \qw		& \qw			& \targ{}	& \qw q_4   	& \qw		& \qw		& \qw		& \qw q_4	& \qw		& \qw		& \qw		& \targ{}      	& \qw q_4	& \qw				& \qw			 	
		\end{quantikz}
	\end{adjustbox}
	\caption{%
		Implementation of Dicke state $\dicke{6}{3}$ on LNN starting from an interleaved ordering of the qubit blocks $(q_1,q_2,q_3)$ and $(q_4,q_5,q_6)$:
		We use the interjacently placed qubits $q_6,q_5,q_4$ while preparing the qubits $q_1,q_2,q_3$. While doing so, their assignment is flipped 
		compared to the supposed superposition, hence we apply $X$-gates in the last step of the ``Divide'' part. 
		$CR_y$ gates still target qubits in a classical state, hence they can be implemented with 1 {\CNOT} each.
		Un-interleaving now needs {\SWAP}s to be decomposed into 3 {\CNOT}s each. Furthermore, the circuit parallelizes less than 
		the method given in Figure~\ref{fig:divide-LNN-D63-33}.\newline 
		For the ``Divide'' part we get a {\CNOT} count of 14 and a {\CNOT} depth of 11, while for the ``Conquer'' part in the big/little-endian unitaries
		$U_{3,3}^{be},\ U_{3,3}^{le}$ we have 9 {\CNOT}s. Overall we get a {\CNOT} count of 32 and a {\CNOT} depth of 20.
	}
	\label{fig:divide-LNN-D63-32}
\end{figure*}

\subsection{``Divide-and-Conquer'' for LNN Topologies}

In our experiments, we are restricted to way sparser architectures. 
Hence, we additionally fine-tune the ``Divide'' techniques for Ladder architectures to Linear Nearest Neighbor architectures. We utilized two methods for this:

In the simplest way, we start from the Ladder architecture of Figure~\ref{fig:divide-conquer-ladder} and remove vertical connectivities except for $(q_1,q_{2n_1})$.
We then exchange for each qubit $q_i,\ i<n_1$ the order of the $\mathit{CR}_y$ and the subtraction {\CNOT} it controls. 
When adding {\SWAP}s and combining each $\mathit{CR}_y$ gate with one of the {\SWAP} gates,
we get the LNN circuit in Figure~\ref{fig:divide-LNN-D63-33}.
Note that, in this case, the {\SWAP}s as well as the combination of $\mathit{CR}_y$ and {\SWAP} gates compile to two {\CNOT}s only, since at the moment of applying the gates, their target is still in a classical state.

An additional option is given in Figure~\ref{fig:divide-LNN-D63-32}. Here the circuits starts with the 
qubits $(q_1,\ldots,q_{n_1})$ and $(q_n,\ldots,q_{n_1+1})$ interleaved. This approach works best for the Dicke state $\dicke{6}{2}$ and also decreased the {\CNOT} count for the Dicke state $\dicke{6}{3}$ while simultaneously increasing the {\CNOT} depth. 
Hence we used this approach in experiments for exclusively for $\dicke{6}{2}$ and both approaches for $\dicke{6}{3}$.

For $k=1$ the first approach retrieves the circuits of previous W-state experiments~\cite{epfl2019wstate}. 
Interactive drag-and-drop implementations of our circuits in the Quirk simulator~\cite{quirk} are behind the following links:
\href{https://algassert.com/quirk#circuit=%7B%22cols%22%3A%5B%5B%22X%22%2C%7B%22id%22%3A%22Ryft%22%2C%22arg%22%3A%222acos(sqrt(1%2F2))%22%7D%5D%2C%5B%5D%2C%5B%5D%2C%5B%5D%2C%5B%22X%22%2C%22%E2%80%A2%22%5D%5D%7D
}{$\dicke{2}{1}$}, 
\href{https://algassert.com/quirk#circuit=%7B%22cols%22%3A%5B%5B%22X%22%2C%7B%22id%22%3A%22Ryft%22%2C%22arg%22%3A%222acos(sqrt(1%2F3))%22%7D%2C%7B%22id%22%3A%22Ryft%22%2C%22arg%22%3A%22acos(sqrt(1%2F2))%22%7D%5D%2C%5B%5D%2C%5B%5D%2C%5B%5D%2C%5B%22X%22%2C%22%E2%80%A2%22%5D%2C%5B1%2C%22%E2%80%A2%22%2C%22X%22%5D%2C%5B1%2C1%2C%7B%22id%22%3A%22Ryft%22%2C%22arg%22%3A%22-acos(sqrt(1%2F2))%22%7D%5D%2C%5B%5D%2C%5B%5D%2C%5B%5D%2C%5B1%2C%22X%22%2C%22%E2%80%A2%22%5D%5D%7D
}{$\dicke{3}{1}$}, 
\href{https://algassert.com/quirk#circuit=%7B%22cols%22%3A%5B%5B%7B%22id%22%3A%22Ryft%22%2C%22arg%22%3A%22acos(sqrt(1%2F2))%22%7D%2C%22X%22%2C%7B%22id%22%3A%22Ryft%22%2C%22arg%22%3A%222acos(sqrt(1%2F2))%22%7D%2C%7B%22id%22%3A%22Ryft%22%2C%22arg%22%3A%22acos(sqrt(1%2F2))%22%7D%5D%2C%5B%5D%2C%5B%5D%2C%5B%5D%2C%5B1%2C%22X%22%2C%22%E2%80%A2%22%5D%2C%5B%22X%22%2C%22%E2%80%A2%22%5D%2C%5B1%2C1%2C%22%E2%80%A2%22%2C%22X%22%5D%2C%5B%7B%22id%22%3A%22Ryft%22%2C%22arg%22%3A%22-acos(sqrt(1%2F2))%22%7D%2C1%2C1%2C%7B%22id%22%3A%22Ryft%22%2C%22arg%22%3A%22-acos(sqrt(1%2F2))%22%7D%5D%2C%5B%5D%2C%5B%5D%2C%5B%5D%2C%5B%22%E2%80%A2%22%2C%22X%22%5D%2C%5B1%2C1%2C%22X%22%2C%22%E2%80%A2%22%5D%5D%7D
}{$\dicke{4}{1}$}, 
\href{https://algassert.com/quirk#circuit=%7B%22cols%22%3A%5B%5B%7B%22id%22%3A%22Ryft%22%2C%22arg%22%3A%22acos(sqrt(1%2F5))%22%7D%2C%7B%22id%22%3A%22Ryft%22%2C%22arg%22%3A%222acos(sqrt(1%2F6))%22%7D%5D%2C%5B%5D%2C%5B%5D%2C%5B%5D%2C%5B1%2C%22%E2%80%A2%22%2C%22X%22%5D%2C%5B1%2C1%2C%22%E2%80%A2%22%2C%22X%22%5D%2C%5B1%2C1%2C%22X%22%2C%22%E2%80%A2%22%5D%2C%5B%22%E2%80%A2%22%2C%22X%22%5D%2C%5B%22X%22%2C%22%E2%80%A2%22%5D%2C%5B1%2C%7B%22id%22%3A%22Ryft%22%2C%22arg%22%3A%22-acos(sqrt(1%2F5))%22%7D%2C%22X%22%2C%22X%22%5D%2C%5B%5D%2C%5B%5D%2C%5B%5D%2C%5B1%2C%22%E2%80%A2%22%2C%22X%22%5D%2C%5B%22Amps4%22%5D%2C%5B%5D%2C%5B%5D%2C%5B%5D%2C%5B%22Chance2%22%2C1%2C%22Chance2%22%5D%2C%5B1%2C%7B%22id%22%3A%22Ryft%22%2C%22arg%22%3A%22pi%2F2%22%7D%2C1%2C%7B%22id%22%3A%22Ryft%22%2C%22arg%22%3A%22pi%2F2%22%7D%5D%2C%5B%22X%22%2C%22%E2%80%A2%22%5D%2C%5B1%2C1%2C%22X%22%2C%22%E2%80%A2%22%5D%2C%5B%7B%22id%22%3A%22Ryft%22%2C%22arg%22%3A%22acos(sqrt(1%2F2))%22%7D%2C%7B%22id%22%3A%22Ryft%22%2C%22arg%22%3A%22acos(sqrt(1%2F2))%22%7D%2C%7B%22id%22%3A%22Ryft%22%2C%22arg%22%3A%22acos(sqrt(1%2F2))%22%7D%2C%7B%22id%22%3A%22Ryft%22%2C%22arg%22%3A%22acos(sqrt(1%2F2))%22%7D%5D%2C%5B%5D%2C%5B%5D%2C%5B%5D%2C%5B%22X%22%2C%22%E2%80%A2%22%5D%2C%5B1%2C1%2C%22X%22%2C%22%E2%80%A2%22%5D%2C%5B1%2C%7B%22id%22%3A%22Ryft%22%2C%22arg%22%3A%22-pi%2F2%22%7D%2C1%2C%7B%22id%22%3A%22Ryft%22%2C%22arg%22%3A%22-pi%2F2%22%7D%5D%5D%7D
}{$\dicke{4}{2}$}, 
\href{https://algassert.com/quirk#circuit=%7B%22cols%22%3A%5B%5B%7B%22id%22%3A%22Ryft%22%2C%22arg%22%3A%22acos(sqrt(1%2F2))%22%7D%2C%22X%22%2C%7B%22id%22%3A%22Ryft%22%2C%22arg%22%3A%222acos(sqrt(2%2F5))%22%7D%2C%7B%22id%22%3A%22Ryft%22%2C%22arg%22%3A%22acos(sqrt(2%2F3))%22%7D%2C%7B%22id%22%3A%22Ryft%22%2C%22arg%22%3A%22acos(sqrt(1%2F2))%22%7D%5D%2C%5B%5D%2C%5B%5D%2C%5B%5D%2C%5B1%2C%22X%22%2C%22%E2%80%A2%22%5D%2C%5B1%2C1%2C%22%E2%80%A2%22%2C%22X%22%5D%2C%5B1%2C1%2C1%2C%7B%22id%22%3A%22Ryft%22%2C%22arg%22%3A%22-acos(sqrt(2%2F3))%22%7D%5D%2C%5B%5D%2C%5B%5D%2C%5B%5D%2C%5B1%2C1%2C%22X%22%2C%22%E2%80%A2%22%5D%2C%5B%22X%22%2C%22%E2%80%A2%22%5D%2C%5B1%2C1%2C1%2C%22%E2%80%A2%22%2C%22X%22%5D%2C%5B%7B%22id%22%3A%22Ryft%22%2C%22arg%22%3A%22-acos(sqrt(1%2F2))%22%7D%2C1%2C1%2C1%2C%7B%22id%22%3A%22Ryft%22%2C%22arg%22%3A%22-acos(sqrt(1%2F2))%22%7D%5D%2C%5B%5D%2C%5B%5D%2C%5B%5D%2C%5B%22%E2%80%A2%22%2C%22X%22%5D%2C%5B1%2C1%2C1%2C%22X%22%2C%22%E2%80%A2%22%5D%5D%7D
}{$\dicke{5}{1}$}, 
\href{https://algassert.com/quirk#circuit=%7B%22cols%22%3A%5B%5B%7B%22id%22%3A%22Ryft%22%2C%22arg%22%3A%22acos(sqrt(1%2F7))%22%7D%2C%7B%22id%22%3A%22Ryft%22%2C%22arg%22%3A%222acos(sqrt(3%2F10))%22%7D%5D%2C%5B%5D%2C%5B%5D%2C%5B%5D%2C%5B1%2C%22%E2%80%A2%22%2C%22X%22%5D%2C%5B1%2C1%2C%22%E2%80%A2%22%2C%22X%22%5D%2C%5B1%2C1%2C%22X%22%2C%22%E2%80%A2%22%5D%2C%5B%22%E2%80%A2%22%2C%22X%22%5D%2C%5B%22X%22%2C%22%E2%80%A2%22%5D%2C%5B1%2C%7B%22id%22%3A%22Ryft%22%2C%22arg%22%3A%22-acos(sqrt(1%2F7))%22%7D%2C%22X%22%2C%22X%22%5D%2C%5B%5D%2C%5B%5D%2C%5B%5D%2C%5B1%2C%22%E2%80%A2%22%2C%22X%22%5D%2C%5B%22Amps4%22%5D%2C%5B%5D%2C%5B%5D%2C%5B%5D%2C%5B%22Chance2%22%2C1%2C%22Chance2%22%5D%2C%5B1%2C1%2C1%2C%7B%22id%22%3A%22Ryft%22%2C%22arg%22%3A%22pi%2F2%22%7D%5D%2C%5B1%2C1%2C%22X%22%2C%22%E2%80%A2%22%5D%2C%5B1%2C1%2C%7B%22id%22%3A%22Ryft%22%2C%22arg%22%3A%22acos(sqrt(2%2F3))%22%7D%2C%7B%22id%22%3A%22Ryft%22%2C%22arg%22%3A%22acos(sqrt(2%2F3))%22%7D%5D%2C%5B%5D%2C%5B%5D%2C%5B%5D%2C%5B1%2C1%2C%22X%22%2C%22%E2%80%A2%22%5D%2C%5B1%2C1%2C1%2C%7B%22id%22%3A%22Ryft%22%2C%22arg%22%3A%22-pi%2F2%22%7D%5D%2C%5B%5D%2C%5B1%2C1%2C1%2C%22%E2%80%A2%22%2C%22X%22%5D%2C%5B1%2C1%2C1%2C%7B%22id%22%3A%22Ryft%22%2C%22arg%22%3A%22-0.5*acos(sqrt(2%2F3))%22%7D%5D%2C%5B%5D%2C%5B%5D%2C%5B%5D%2C%5B%5D%2C%5B1%2C1%2C%22%E2%80%A2%22%2C%22X%22%5D%2C%5B1%2C1%2C1%2C%7B%22id%22%3A%22Ryft%22%2C%22arg%22%3A%220.5*acos(sqrt(2%2F3))%22%7D%5D%2C%5B%5D%2C%5B%5D%2C%5B%5D%2C%5B1%2C1%2C1%2C%22X%22%2C%22%E2%80%A2%22%5D%2C%5B1%2C1%2C1%2C%7B%22id%22%3A%22Ryft%22%2C%22arg%22%3A%22-0.5*acos(sqrt(2%2F3))%22%7D%5D%2C%5B%5D%2C%5B%5D%2C%5B%5D%2C%5B%5D%2C%5B1%2C1%2C%22%E2%80%A2%22%2C%22X%22%5D%2C%5B1%2C1%2C1%2C%7B%22id%22%3A%22Ryft%22%2C%22arg%22%3A%220.5*acos(sqrt(2%2F3))%22%7D%5D%2C%5B%5D%2C%5B%5D%2C%5B%5D%2C%5B1%2C1%2C1%2C%22%E2%80%A2%22%2C%22X%22%5D%2C%5B1%2C%7B%22id%22%3A%22Ryft%22%2C%22arg%22%3A%22pi%2F2%22%7D%2C1%2C%7B%22id%22%3A%22Ryft%22%2C%22arg%22%3A%22pi%2F2%22%7D%5D%2C%5B%22X%22%2C%22%E2%80%A2%22%5D%2C%5B1%2C1%2C%22X%22%2C%22%E2%80%A2%22%5D%2C%5B%7B%22id%22%3A%22Ryft%22%2C%22arg%22%3A%22acos(sqrt(1%2F2))%22%7D%2C%7B%22id%22%3A%22Ryft%22%2C%22arg%22%3A%22acos(sqrt(1%2F2))%22%7D%2C%7B%22id%22%3A%22Ryft%22%2C%22arg%22%3A%22acos(sqrt(1%2F2))%22%7D%2C%7B%22id%22%3A%22Ryft%22%2C%22arg%22%3A%22acos(sqrt(1%2F2))%22%7D%5D%2C%5B%5D%2C%5B%5D%2C%5B%5D%2C%5B%22X%22%2C%22%E2%80%A2%22%5D%2C%5B1%2C1%2C%22X%22%2C%22%E2%80%A2%22%5D%2C%5B1%2C%7B%22id%22%3A%22Ryft%22%2C%22arg%22%3A%22-pi%2F2%22%7D%2C1%2C%7B%22id%22%3A%22Ryft%22%2C%22arg%22%3A%22-pi%2F2%22%7D%5D%5D%7D
}{$\dicke{5}{2}$}, 
\href{https://algassert.com/quirk#circuit=%7B%22cols%22%3A%5B%5B%7B%22id%22%3A%22Ryft%22%2C%22arg%22%3A%22acos(sqrt(1%2F2))%22%7D%2C%7B%22id%22%3A%22Ryft%22%2C%22arg%22%3A%22acos(sqrt(2%2F3))%22%7D%2C%22X%22%2C%7B%22id%22%3A%22Ryft%22%2C%22arg%22%3A%222acos(sqrt(1%2F2))%22%7D%2C%7B%22id%22%3A%22Ryft%22%2C%22arg%22%3A%22acos(sqrt(2%2F3))%22%7D%2C%7B%22id%22%3A%22Ryft%22%2C%22arg%22%3A%22acos(sqrt(1%2F2))%22%7D%5D%2C%5B%5D%2C%5B%5D%2C%5B%5D%2C%5B1%2C1%2C%22X%22%2C%22%E2%80%A2%22%5D%2C%5B1%2C%22X%22%2C%22%E2%80%A2%22%5D%2C%5B1%2C1%2C1%2C%22%E2%80%A2%22%2C%22X%22%5D%2C%5B1%2C%7B%22id%22%3A%22Ryft%22%2C%22arg%22%3A%22-acos(sqrt(2%2F3))%22%7D%2C1%2C1%2C%7B%22id%22%3A%22Ryft%22%2C%22arg%22%3A%22-acos(sqrt(2%2F3))%22%7D%5D%2C%5B%5D%2C%5B%5D%2C%5B%5D%2C%5B1%2C%22%E2%80%A2%22%2C%22X%22%5D%2C%5B1%2C1%2C1%2C%22X%22%2C%22%E2%80%A2%22%5D%2C%5B%22X%22%2C%22%E2%80%A2%22%5D%2C%5B1%2C1%2C1%2C1%2C%22%E2%80%A2%22%2C%22X%22%5D%2C%5B%7B%22id%22%3A%22Ryft%22%2C%22arg%22%3A%22-acos(sqrt(1%2F2))%22%7D%2C1%2C1%2C1%2C1%2C%7B%22id%22%3A%22Ryft%22%2C%22arg%22%3A%22-acos(sqrt(1%2F2))%22%7D%5D%2C%5B%5D%2C%5B%5D%2C%5B%5D%2C%5B%22%E2%80%A2%22%2C%22X%22%5D%2C%5B1%2C1%2C1%2C1%2C%22X%22%2C%22%E2%80%A2%22%5D%5D%7D
}{$\dicke{6}{1}$}, 
\href{https://algassert.com/quirk#circuit=%7B%22cols%22%3A%5B%5B1%2C%7B%22id%22%3A%22Ryft%22%2C%22arg%22%3A%222acos(sqrt(3%2F15))%22%7D%2C1%2C%7B%22id%22%3A%22Ryft%22%2C%22arg%22%3A%22acos(sqrt(3%2F12))%22%7D%5D%2C%5B%5D%2C%5B%5D%2C%5B%5D%2C%5B1%2C%22%E2%80%A2%22%2C%22X%22%5D%2C%5B1%2C1%2C%22%E2%80%A2%22%2C%22X%22%5D%2C%5B1%2C1%2C1%2C%7B%22id%22%3A%22Ryft%22%2C%22arg%22%3A%22-acos(sqrt(3%2F12))%22%7D%5D%2C%5B%5D%2C%5B%5D%2C%5B%5D%2C%5B1%2C1%2C1%2C%22%E2%80%A2%22%2C%22X%22%5D%2C%5B1%2C1%2C%22X%22%2C%22%E2%80%A2%22%5D%2C%5B1%2C1%2C%22%E2%80%A2%22%2C%22X%22%5D%2C%5B1%2C1%2C%22X%22%2C%22%E2%80%A2%22%5D%2C%5B1%2C%22X%22%2C%22X%22%5D%2C%5B1%2C%22Amps4%22%5D%2C%5B%5D%2C%5B%5D%2C%5B%5D%2C%5B1%2C%22Chance2%22%2C1%2C%22Chance2%22%5D%2C%5B1%2C%7B%22id%22%3A%22Ryft%22%2C%22arg%22%3A%22pi%2F2%22%7D%2C1%2C1%2C%7B%22id%22%3A%22Ryft%22%2C%22arg%22%3A%22pi%2F2%22%7D%5D%2C%5B1%2C%22%E2%80%A2%22%2C%22X%22%5D%2C%5B1%2C1%2C1%2C%22X%22%2C%22%E2%80%A2%22%5D%2C%5B1%2C%7B%22id%22%3A%22Ryft%22%2C%22arg%22%3A%22acos(sqrt(2%2F3))%22%7D%2C%7B%22id%22%3A%22Ryft%22%2C%22arg%22%3A%22acos(sqrt(2%2F3))%22%7D%2C%7B%22id%22%3A%22Ryft%22%2C%22arg%22%3A%22acos(sqrt(2%2F3))%22%7D%2C%7B%22id%22%3A%22Ryft%22%2C%22arg%22%3A%22acos(sqrt(2%2F3))%22%7D%5D%2C%5B%5D%2C%5B%5D%2C%5B%5D%2C%5B1%2C%22%E2%80%A2%22%2C%22X%22%5D%2C%5B1%2C1%2C1%2C%22X%22%2C%22%E2%80%A2%22%5D%2C%5B1%2C%7B%22id%22%3A%22Ryft%22%2C%22arg%22%3A%22-pi%2F2%22%7D%2C1%2C1%2C%7B%22id%22%3A%22Ryft%22%2C%22arg%22%3A%22-pi%2F2%22%7D%5D%2C%5B%5D%2C%5B%22X%22%2C%22%E2%80%A2%22%5D%2C%5B1%2C1%2C1%2C1%2C%22%E2%80%A2%22%2C%22X%22%5D%2C%5B1%2C%7B%22id%22%3A%22Ryft%22%2C%22arg%22%3A%22-0.5*acos(sqrt(2%2F3))%22%7D%2C1%2C1%2C%7B%22id%22%3A%22Ryft%22%2C%22arg%22%3A%22-0.5*acos(sqrt(2%2F3))%22%7D%5D%2C%5B%5D%2C%5B%5D%2C%5B%5D%2C%5B%5D%2C%5B1%2C%22X%22%2C%22%E2%80%A2%22%5D%2C%5B1%2C1%2C1%2C%22%E2%80%A2%22%2C%22X%22%5D%2C%5B1%2C%7B%22id%22%3A%22Ryft%22%2C%22arg%22%3A%220.5*acos(sqrt(2%2F3))%22%7D%2C1%2C1%2C%7B%22id%22%3A%22Ryft%22%2C%22arg%22%3A%220.5*acos(sqrt(2%2F3))%22%7D%5D%2C%5B%5D%2C%5B%5D%2C%5B%5D%2C%5B%22%E2%80%A2%22%2C%22X%22%5D%2C%5B1%2C1%2C1%2C1%2C%22X%22%2C%22%E2%80%A2%22%5D%2C%5B1%2C%7B%22id%22%3A%22Ryft%22%2C%22arg%22%3A%22-0.5*acos(sqrt(2%2F3))%22%7D%2C1%2C1%2C%7B%22id%22%3A%22Ryft%22%2C%22arg%22%3A%22-0.5*acos(sqrt(2%2F3))%22%7D%5D%2C%5B%5D%2C%5B%5D%2C%5B%5D%2C%5B%5D%2C%5B1%2C%22X%22%2C%22%E2%80%A2%22%5D%2C%5B1%2C1%2C1%2C%22%E2%80%A2%22%2C%22X%22%5D%2C%5B1%2C%7B%22id%22%3A%22Ryft%22%2C%22arg%22%3A%220.5*acos(sqrt(2%2F3))%22%7D%2C1%2C1%2C%7B%22id%22%3A%22Ryft%22%2C%22arg%22%3A%220.5*acos(sqrt(2%2F3))%22%7D%5D%2C%5B%5D%2C%5B%5D%2C%5B%5D%2C%5B%22X%22%2C%22%E2%80%A2%22%5D%2C%5B1%2C1%2C1%2C1%2C%22%E2%80%A2%22%2C%22X%22%5D%2C%5B1%2C%7B%22id%22%3A%22Ryft%22%2C%22arg%22%3A%22pi%2F2%22%7D%2C1%2C1%2C%7B%22id%22%3A%22Ryft%22%2C%22arg%22%3A%22pi%2F2%22%7D%5D%2C%5B1%2C%22%E2%80%A2%22%2C%22X%22%5D%2C%5B1%2C1%2C1%2C%22X%22%2C%22%E2%80%A2%22%5D%2C%5B1%2C%7B%22id%22%3A%22Ryft%22%2C%22arg%22%3A%22acos(sqrt(1%2F2))%22%7D%2C%7B%22id%22%3A%22Ryft%22%2C%22arg%22%3A%22acos(sqrt(1%2F2))%22%7D%2C%7B%22id%22%3A%22Ryft%22%2C%22arg%22%3A%22acos(sqrt(1%2F2))%22%7D%2C%7B%22id%22%3A%22Ryft%22%2C%22arg%22%3A%22acos(sqrt(1%2F2))%22%7D%5D%2C%5B%5D%2C%5B%5D%2C%5B%5D%2C%5B1%2C%22%E2%80%A2%22%2C%22X%22%5D%2C%5B1%2C1%2C1%2C%22X%22%2C%22%E2%80%A2%22%5D%2C%5B1%2C%7B%22id%22%3A%22Ryft%22%2C%22arg%22%3A%22-pi%2F2%22%7D%2C1%2C1%2C%7B%22id%22%3A%22Ryft%22%2C%22arg%22%3A%22-pi%2F2%22%7D%5D%5D%7D
}{$\dicke{6}{2}$ (interleaved)}, 
\href{https://algassert.com/quirk#circuit=%7B%22cols%22%3A%5B%5B1%2C%7B%22id%22%3A%22Ryft%22%2C%22arg%22%3A%22acos(sqrt(10%2F19))%22%7D%2C%7B%22id%22%3A%22Ryft%22%2C%22arg%22%3A%222acos(sqrt(1%2F20))%22%7D%2C1%2C%22X%22%2C%22X%22%5D%2C%5B%5D%2C%5B%5D%2C%5B%5D%2C%5B1%2C1%2C%22%E2%80%A2%22%2C%22X%22%5D%2C%5B1%2C1%2C1%2C%22%E2%80%A2%22%2C%22X%22%5D%2C%5B1%2C1%2C1%2C%22X%22%2C%22%E2%80%A2%22%5D%2C%5B1%2C1%2C1%2C1%2C%22X%22%5D%2C%5B1%2C1%2C1%2C1%2C%22%E2%80%A2%22%2C%22X%22%5D%2C%5B1%2C1%2C1%2C1%2C%22X%22%2C%22%E2%80%A2%22%5D%2C%5B1%2C%22%E2%80%A2%22%2C%22X%22%5D%2C%5B1%2C%22X%22%2C%22%E2%80%A2%22%5D%2C%5B1%2C1%2C%7B%22id%22%3A%22Ryft%22%2C%22arg%22%3A%22-acos(sqrt(10%2F19))%22%7D%5D%2C%5B%5D%2C%5B%5D%2C%5B%5D%2C%5B%22X%22%2C%22%E2%80%A2%22%5D%2C%5B%22%E2%80%A2%22%2C%22X%22%5D%2C%5B1%2C1%2C%22%E2%80%A2%22%2C%22X%22%5D%2C%5B1%2C1%2C1%2C%22X%22%5D%2C%5B1%2C1%2C1%2C%22%E2%80%A2%22%2C%22X%22%5D%2C%5B1%2C1%2C1%2C%22X%22%2C%22%E2%80%A2%22%5D%2C%5B1%2C%7B%22id%22%3A%22Ryft%22%2C%22arg%22%3A%22acos(sqrt(1%2F10))%22%7D%5D%2C%5B%5D%2C%5B%5D%2C%5B%5D%2C%5B1%2C%22%E2%80%A2%22%2C%22X%22%5D%2C%5B1%2C%22X%22%2C%22%E2%80%A2%22%5D%2C%5B1%2C1%2C%7B%22id%22%3A%22Ryft%22%2C%22arg%22%3A%22-acos(sqrt(1%2F10))%22%7D%5D%2C%5B%5D%2C%5B%5D%2C%5B%5D%2C%5B1%2C1%2C%22%E2%80%A2%22%2C%22X%22%5D%2C%5B%22Amps6%22%5D%2C%5B%5D%2C%5B%5D%2C%5B%5D%2C%5B%5D%2C%5B%5D%2C%5B%22Chance3%22%2C1%2C1%2C%22Chance3%22%5D%2C%5B1%2C%7B%22id%22%3A%22Ryft%22%2C%22arg%22%3A%22pi%2F2%22%7D%2C1%2C1%2C%7B%22id%22%3A%22Ryft%22%2C%22arg%22%3A%22pi%2F2%22%7D%5D%2C%5B%22X%22%2C%22%E2%80%A2%22%5D%2C%5B1%2C1%2C1%2C%22X%22%2C%22%E2%80%A2%22%5D%2C%5B%7B%22id%22%3A%22Ryft%22%2C%22arg%22%3A%22acos(sqrt(2%2F3))%22%7D%2C%7B%22id%22%3A%22Ryft%22%2C%22arg%22%3A%22acos(sqrt(2%2F3))%22%7D%2C1%2C%7B%22id%22%3A%22Ryft%22%2C%22arg%22%3A%22acos(sqrt(2%2F3))%22%7D%2C%7B%22id%22%3A%22Ryft%22%2C%22arg%22%3A%22acos(sqrt(2%2F3))%22%7D%5D%2C%5B%5D%2C%5B%5D%2C%5B%5D%2C%5B%22X%22%2C%22%E2%80%A2%22%5D%2C%5B1%2C1%2C1%2C%22X%22%2C%22%E2%80%A2%22%5D%2C%5B1%2C%7B%22id%22%3A%22Ryft%22%2C%22arg%22%3A%22-pi%2F2%22%7D%2C1%2C1%2C%7B%22id%22%3A%22Ryft%22%2C%22arg%22%3A%22-pi%2F2%22%7D%5D%2C%5B%5D%2C%5B1%2C%22%E2%80%A2%22%2C%22X%22%5D%2C%5B1%2C1%2C1%2C1%2C%22%E2%80%A2%22%2C%22X%22%5D%2C%5B1%2C%7B%22id%22%3A%22Ryft%22%2C%22arg%22%3A%22-0.5*acos(sqrt(2%2F3))%22%7D%2C1%2C1%2C%7B%22id%22%3A%22Ryft%22%2C%22arg%22%3A%22-0.5*acos(sqrt(2%2F3))%22%7D%5D%2C%5B%5D%2C%5B%5D%2C%5B%5D%2C%5B%5D%2C%5B%22%E2%80%A2%22%2C%22X%22%5D%2C%5B1%2C1%2C1%2C%22%E2%80%A2%22%2C%22X%22%5D%2C%5B1%2C%7B%22id%22%3A%22Ryft%22%2C%22arg%22%3A%220.5*acos(sqrt(2%2F3))%22%7D%2C1%2C1%2C%7B%22id%22%3A%22Ryft%22%2C%22arg%22%3A%220.5*acos(sqrt(2%2F3))%22%7D%5D%2C%5B%5D%2C%5B%5D%2C%5B%5D%2C%5B1%2C%22X%22%2C%22%E2%80%A2%22%5D%2C%5B1%2C1%2C1%2C1%2C%22X%22%2C%22%E2%80%A2%22%5D%2C%5B1%2C%7B%22id%22%3A%22Ryft%22%2C%22arg%22%3A%22-0.5*acos(sqrt(2%2F3))%22%7D%2C1%2C1%2C%7B%22id%22%3A%22Ryft%22%2C%22arg%22%3A%22-0.5*acos(sqrt(2%2F3))%22%7D%5D%2C%5B%5D%2C%5B%5D%2C%5B%5D%2C%5B%5D%2C%5B%22%E2%80%A2%22%2C%22X%22%5D%2C%5B1%2C1%2C1%2C%22%E2%80%A2%22%2C%22X%22%5D%2C%5B1%2C%7B%22id%22%3A%22Ryft%22%2C%22arg%22%3A%220.5*acos(sqrt(2%2F3))%22%7D%2C1%2C1%2C%7B%22id%22%3A%22Ryft%22%2C%22arg%22%3A%220.5*acos(sqrt(2%2F3))%22%7D%5D%2C%5B%5D%2C%5B%5D%2C%5B%5D%2C%5B1%2C%22%E2%80%A2%22%2C%22X%22%5D%2C%5B1%2C1%2C1%2C1%2C%22%E2%80%A2%22%2C%22X%22%5D%2C%5B1%2C%7B%22id%22%3A%22Ryft%22%2C%22arg%22%3A%22pi%2F2%22%7D%2C1%2C1%2C%7B%22id%22%3A%22Ryft%22%2C%22arg%22%3A%22pi%2F2%22%7D%5D%2C%5B%22X%22%2C%22%E2%80%A2%22%5D%2C%5B1%2C1%2C1%2C%22X%22%2C%22%E2%80%A2%22%5D%2C%5B%7B%22id%22%3A%22Ryft%22%2C%22arg%22%3A%22acos(sqrt(1%2F2))%22%7D%2C%7B%22id%22%3A%22Ryft%22%2C%22arg%22%3A%22acos(sqrt(1%2F2))%22%7D%2C1%2C%7B%22id%22%3A%22Ryft%22%2C%22arg%22%3A%22acos(sqrt(1%2F2))%22%7D%2C%7B%22id%22%3A%22Ryft%22%2C%22arg%22%3A%22acos(sqrt(1%2F2))%22%7D%5D%2C%5B%5D%2C%5B%5D%2C%5B%5D%2C%5B%22X%22%2C%22%E2%80%A2%22%5D%2C%5B1%2C1%2C1%2C%22X%22%2C%22%E2%80%A2%22%5D%2C%5B1%2C%7B%22id%22%3A%22Ryft%22%2C%22arg%22%3A%22-pi%2F2%22%7D%2C1%2C1%2C%7B%22id%22%3A%22Ryft%22%2C%22arg%22%3A%22-pi%2F2%22%7D%5D%5D%7D
}{$\dicke{6}{3}$}, 
\href{https://algassert.com/quirk#circuit=%7B%22cols%22%3A%5B%5B%7B%22id%22%3A%22Ryft%22%2C%22arg%22%3A%222acos(sqrt(1%2F20))%22%7D%2C1%2C%7B%22id%22%3A%22Ryft%22%2C%22arg%22%3A%22acos(sqrt(10%2F19))%22%7D%2C1%2C%7B%22id%22%3A%22Ryft%22%2C%22arg%22%3A%22acos(sqrt(1%2F10))%22%7D%5D%2C%5B%5D%2C%5B%5D%2C%5B%5D%2C%5B%22%E2%80%A2%22%2C%22X%22%5D%2C%5B1%2C%22%E2%80%A2%22%2C%22X%22%5D%2C%5B1%2C1%2C%7B%22id%22%3A%22Ryft%22%2C%22arg%22%3A%22-acos(sqrt(10%2F19))%22%7D%5D%2C%5B%5D%2C%5B%5D%2C%5B%5D%2C%5B1%2C1%2C%22%E2%80%A2%22%2C%22X%22%5D%2C%5B1%2C1%2C1%2C%22%E2%80%A2%22%2C%22X%22%5D%2C%5B1%2C1%2C1%2C1%2C%7B%22id%22%3A%22Ryft%22%2C%22arg%22%3A%22-acos(sqrt(1%2F10))%22%7D%5D%2C%5B%5D%2C%5B%5D%2C%5B%5D%2C%5B1%2C1%2C1%2C1%2C%22%E2%80%A2%22%2C%22X%22%5D%2C%5B1%2C%22%E2%80%A2%22%2C%22X%22%5D%2C%5B1%2C%22X%22%2C%22%E2%80%A2%22%5D%2C%5B1%2C%22%E2%80%A2%22%2C%22X%22%5D%2C%5B1%2C1%2C1%2C%22%E2%80%A2%22%2C%22X%22%5D%2C%5B1%2C1%2C1%2C%22X%22%2C%22%E2%80%A2%22%5D%2C%5B1%2C1%2C1%2C%22%E2%80%A2%22%2C%22X%22%5D%2C%5B1%2C1%2C%22%E2%80%A2%22%2C%22X%22%5D%2C%5B1%2C1%2C%22X%22%2C%22%E2%80%A2%22%5D%2C%5B1%2C1%2C%22%E2%80%A2%22%2C%22X%22%5D%2C%5B%22X%22%2C%22X%22%2C%22X%22%5D%2C%5B%22Amps6%22%5D%2C%5B%5D%2C%5B%5D%2C%5B%5D%2C%5B%5D%2C%5B%5D%2C%5B%22Chance3%22%2C1%2C1%2C%22Chance3%22%5D%2C%5B1%2C%7B%22id%22%3A%22Ryft%22%2C%22arg%22%3A%22pi%2F2%22%7D%2C1%2C1%2C%7B%22id%22%3A%22Ryft%22%2C%22arg%22%3A%22pi%2F2%22%7D%5D%2C%5B1%2C%22%E2%80%A2%22%2C%22X%22%5D%2C%5B1%2C1%2C1%2C%22X%22%2C%22%E2%80%A2%22%5D%2C%5B1%2C%7B%22id%22%3A%22Ryft%22%2C%22arg%22%3A%22acos(sqrt(2%2F3))%22%7D%2C%7B%22id%22%3A%22Ryft%22%2C%22arg%22%3A%22acos(sqrt(2%2F3))%22%7D%2C%7B%22id%22%3A%22Ryft%22%2C%22arg%22%3A%22acos(sqrt(2%2F3))%22%7D%2C%7B%22id%22%3A%22Ryft%22%2C%22arg%22%3A%22acos(sqrt(2%2F3))%22%7D%5D%2C%5B%5D%2C%5B%5D%2C%5B%5D%2C%5B1%2C%22%E2%80%A2%22%2C%22X%22%5D%2C%5B1%2C1%2C1%2C%22X%22%2C%22%E2%80%A2%22%5D%2C%5B1%2C%7B%22id%22%3A%22Ryft%22%2C%22arg%22%3A%22-pi%2F2%22%7D%2C1%2C1%2C%7B%22id%22%3A%22Ryft%22%2C%22arg%22%3A%22-pi%2F2%22%7D%5D%2C%5B%5D%2C%5B%22X%22%2C%22%E2%80%A2%22%5D%2C%5B1%2C1%2C1%2C1%2C%22%E2%80%A2%22%2C%22X%22%5D%2C%5B1%2C%7B%22id%22%3A%22Ryft%22%2C%22arg%22%3A%22-0.5*acos(sqrt(2%2F3))%22%7D%2C1%2C1%2C%7B%22id%22%3A%22Ryft%22%2C%22arg%22%3A%22-0.5*acos(sqrt(2%2F3))%22%7D%5D%2C%5B%5D%2C%5B%5D%2C%5B%5D%2C%5B%5D%2C%5B1%2C%22X%22%2C%22%E2%80%A2%22%5D%2C%5B1%2C1%2C1%2C%22%E2%80%A2%22%2C%22X%22%5D%2C%5B1%2C%7B%22id%22%3A%22Ryft%22%2C%22arg%22%3A%220.5*acos(sqrt(2%2F3))%22%7D%2C1%2C1%2C%7B%22id%22%3A%22Ryft%22%2C%22arg%22%3A%220.5*acos(sqrt(2%2F3))%22%7D%5D%2C%5B%5D%2C%5B%5D%2C%5B%5D%2C%5B%22%E2%80%A2%22%2C%22X%22%5D%2C%5B1%2C1%2C1%2C1%2C%22X%22%2C%22%E2%80%A2%22%5D%2C%5B1%2C%7B%22id%22%3A%22Ryft%22%2C%22arg%22%3A%22-0.5*acos(sqrt(2%2F3))%22%7D%2C1%2C1%2C%7B%22id%22%3A%22Ryft%22%2C%22arg%22%3A%22-0.5*acos(sqrt(2%2F3))%22%7D%5D%2C%5B%5D%2C%5B%5D%2C%5B%5D%2C%5B%5D%2C%5B1%2C%22X%22%2C%22%E2%80%A2%22%5D%2C%5B1%2C1%2C1%2C%22%E2%80%A2%22%2C%22X%22%5D%2C%5B1%2C%7B%22id%22%3A%22Ryft%22%2C%22arg%22%3A%220.5*acos(sqrt(2%2F3))%22%7D%2C1%2C1%2C%7B%22id%22%3A%22Ryft%22%2C%22arg%22%3A%220.5*acos(sqrt(2%2F3))%22%7D%5D%2C%5B%5D%2C%5B%5D%2C%5B%5D%2C%5B%22X%22%2C%22%E2%80%A2%22%5D%2C%5B1%2C1%2C1%2C1%2C%22%E2%80%A2%22%2C%22X%22%5D%2C%5B1%2C%7B%22id%22%3A%22Ryft%22%2C%22arg%22%3A%22pi%2F2%22%7D%2C1%2C1%2C%7B%22id%22%3A%22Ryft%22%2C%22arg%22%3A%22pi%2F2%22%7D%5D%2C%5B1%2C%22%E2%80%A2%22%2C%22X%22%5D%2C%5B1%2C1%2C1%2C%22X%22%2C%22%E2%80%A2%22%5D%2C%5B1%2C%7B%22id%22%3A%22Ryft%22%2C%22arg%22%3A%22acos(sqrt(1%2F2))%22%7D%2C%7B%22id%22%3A%22Ryft%22%2C%22arg%22%3A%22acos(sqrt(1%2F2))%22%7D%2C%7B%22id%22%3A%22Ryft%22%2C%22arg%22%3A%22acos(sqrt(1%2F2))%22%7D%2C%7B%22id%22%3A%22Ryft%22%2C%22arg%22%3A%22acos(sqrt(1%2F2))%22%7D%5D%2C%5B%5D%2C%5B%5D%2C%5B%5D%2C%5B1%2C%22%E2%80%A2%22%2C%22X%22%5D%2C%5B1%2C1%2C1%2C%22X%22%2C%22%E2%80%A2%22%5D%2C%5B1%2C%7B%22id%22%3A%22Ryft%22%2C%22arg%22%3A%22-pi%2F2%22%7D%2C1%2C1%2C%7B%22id%22%3A%22Ryft%22%2C%22arg%22%3A%22-pi%2F2%22%7D%5D%5D%7D
}{$\dicke{6}{3}$ (interleaved)}.

\section{Methods}

In this section, we discuss our experimental setup, the evaluated metrics (quantum state fidelity, Hellinger fidelity, and measured success probability),
as well as relations between those and existing benchmarks. 

\begin{figure*}[t!]
    \centering
    \includegraphics[width=0.32\textwidth]{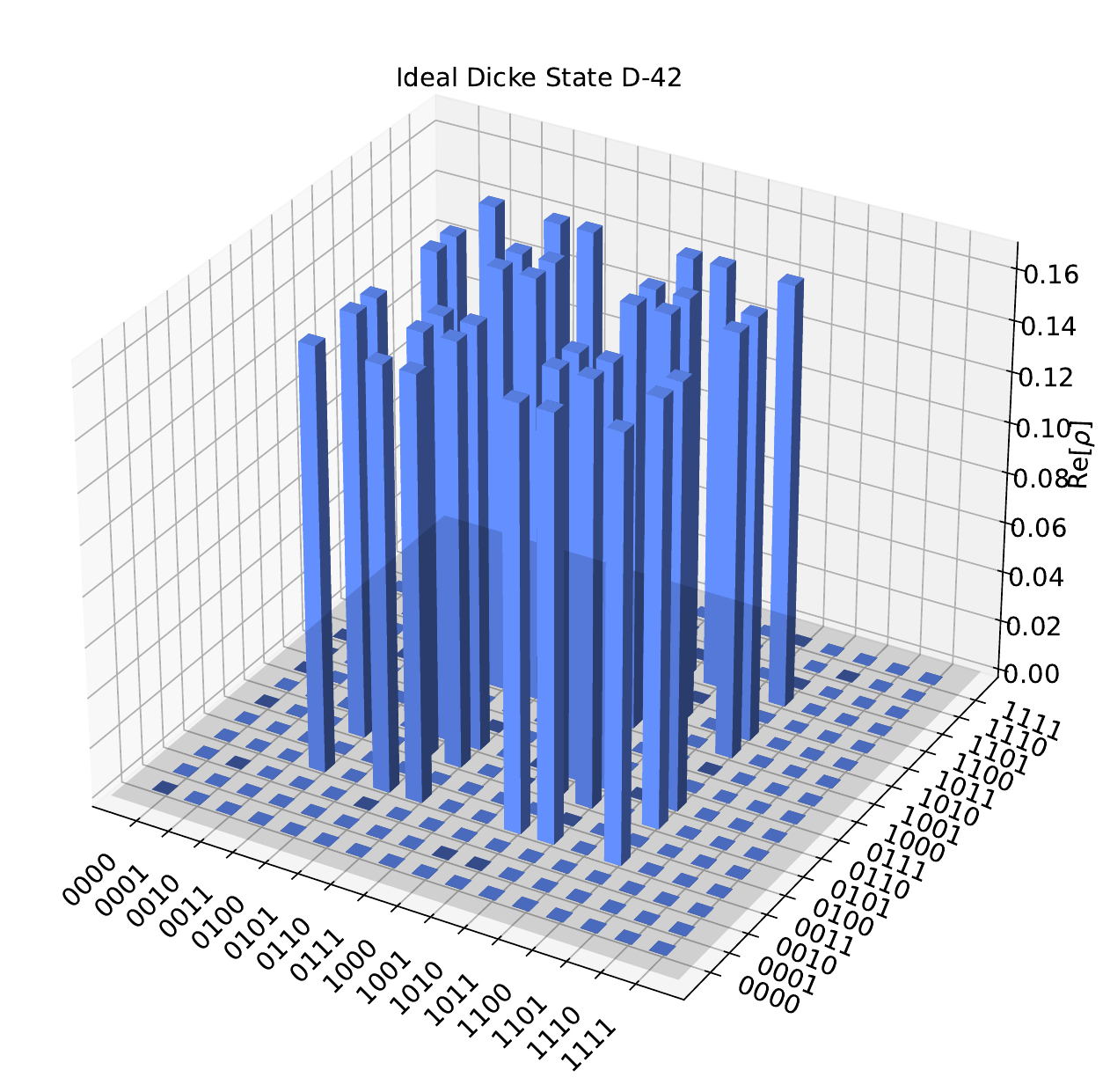}\hfill%
    \includegraphics[width=0.32\textwidth]{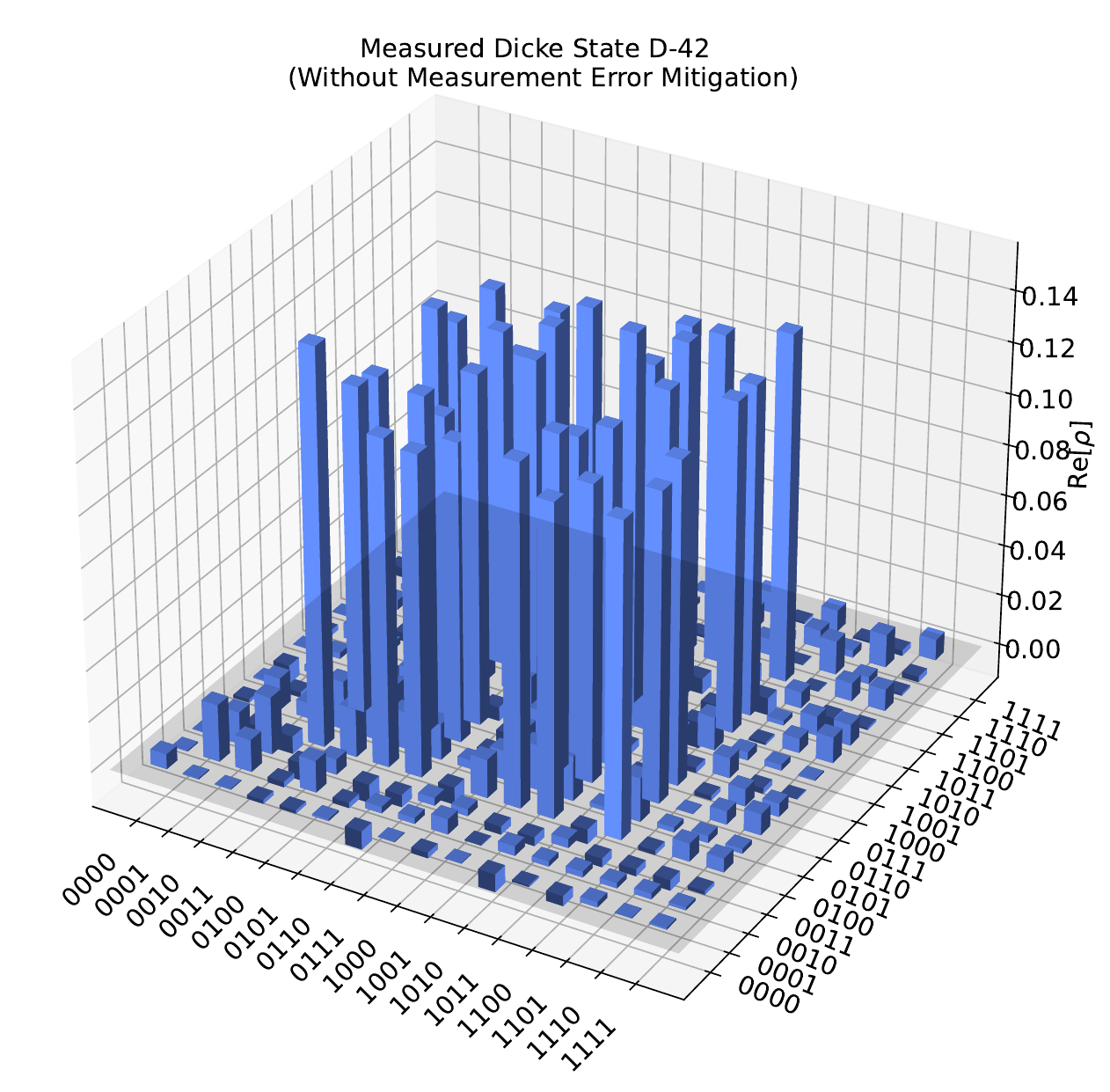}\hfill%
    \includegraphics[width=0.32\textwidth]{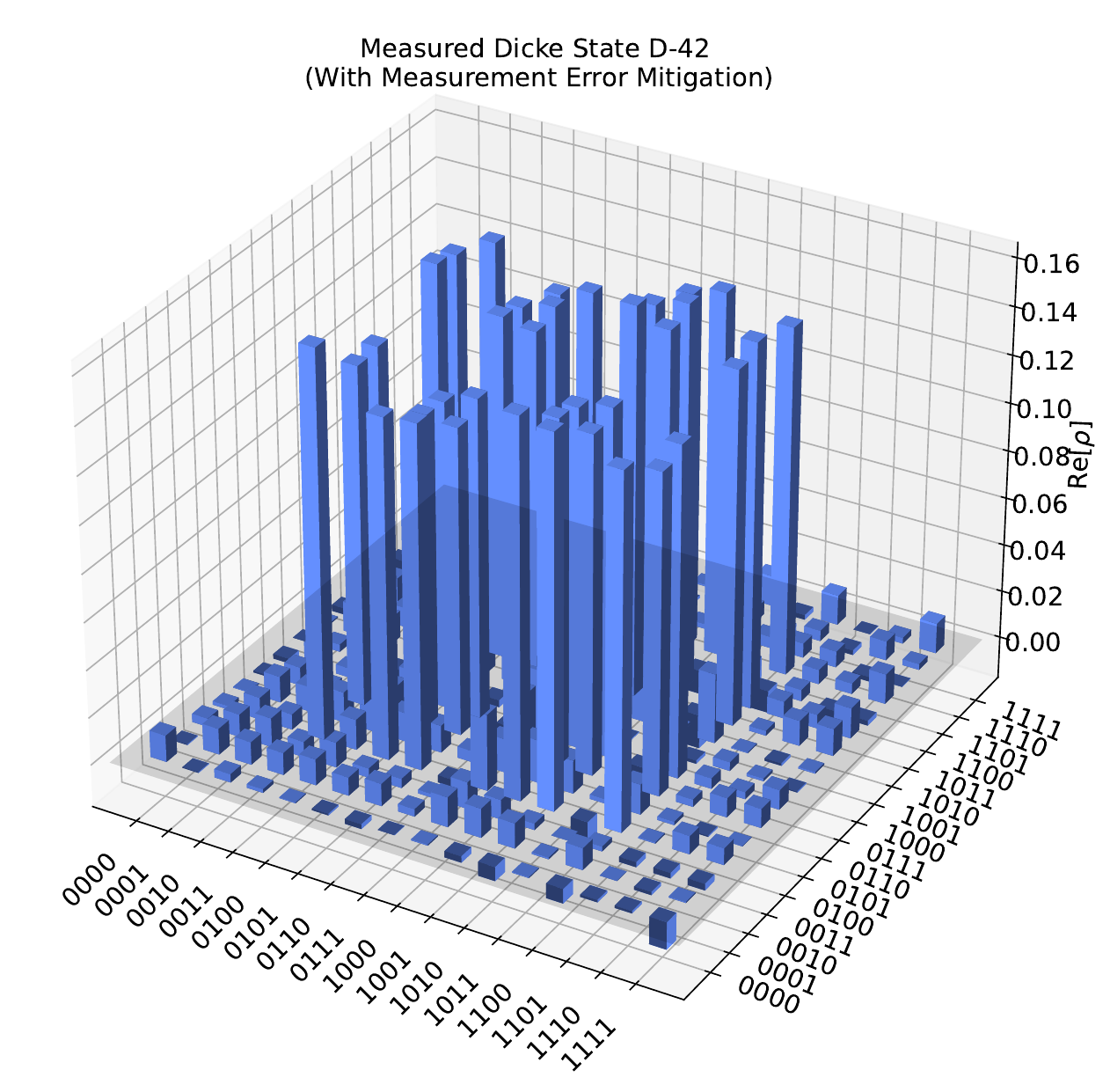}\\
    \includegraphics[width=0.32\textwidth]{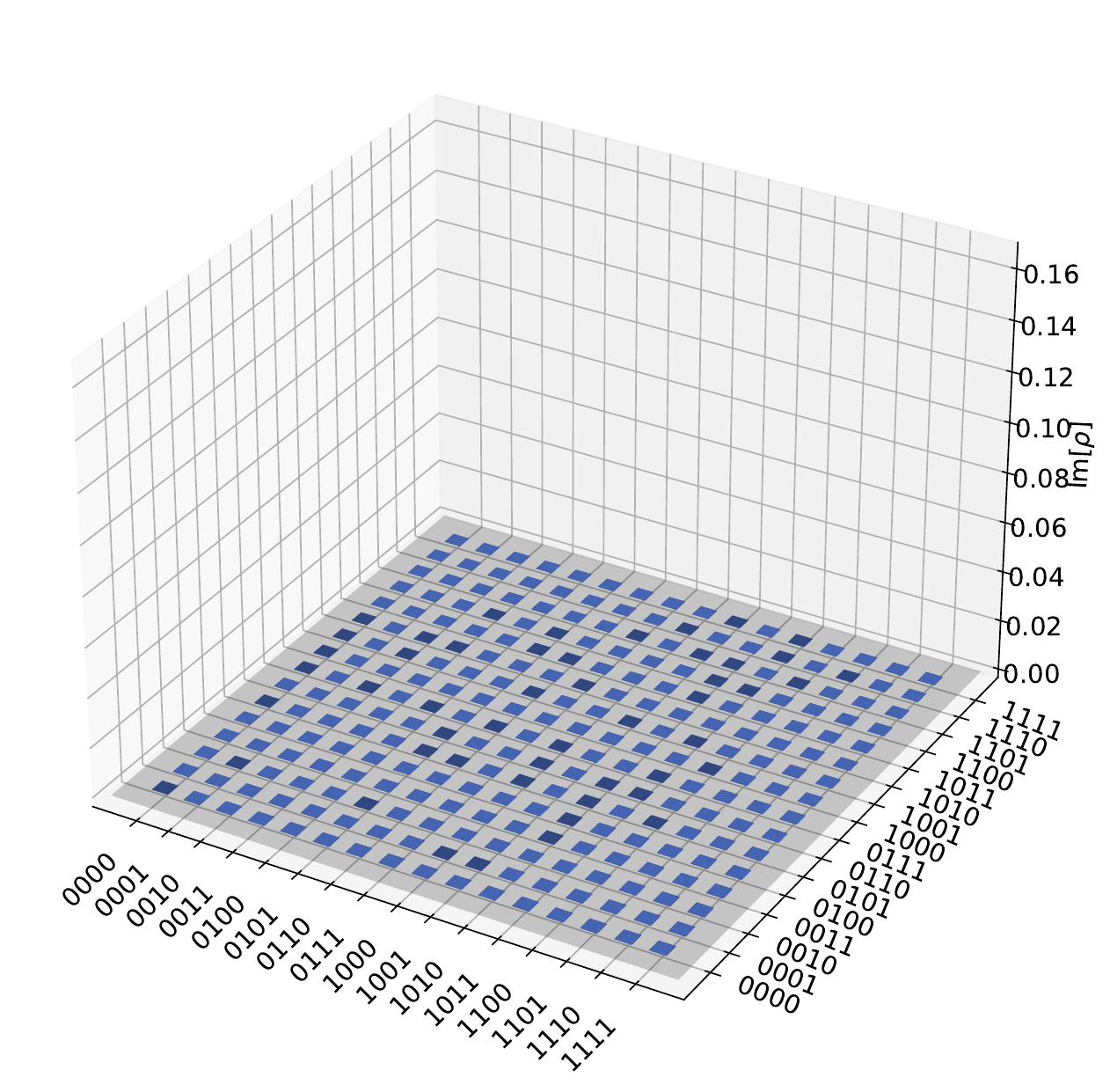}\hfill%
    \includegraphics[width=0.32\textwidth]{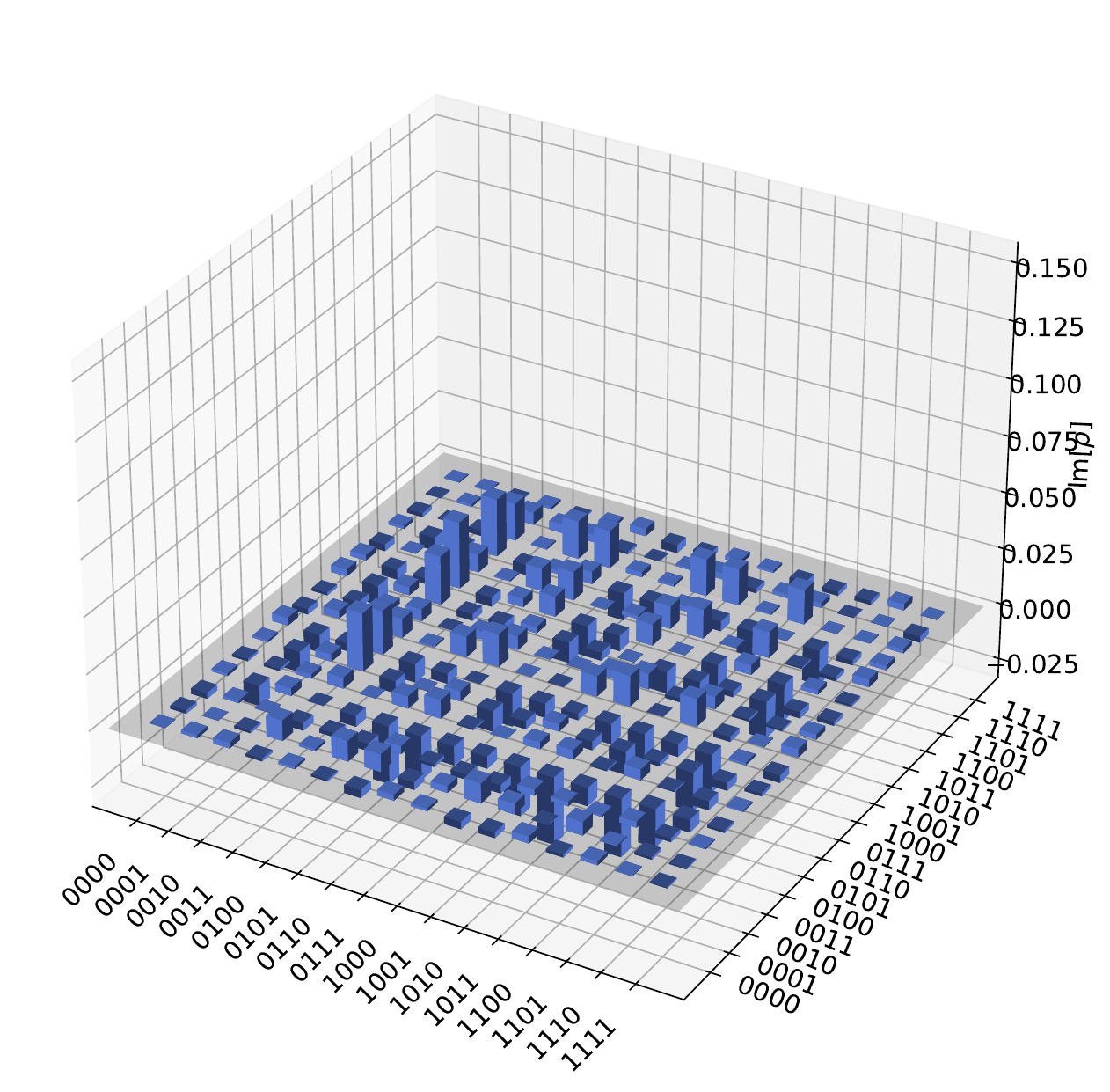}\hfill%
    \includegraphics[width=0.32\textwidth]{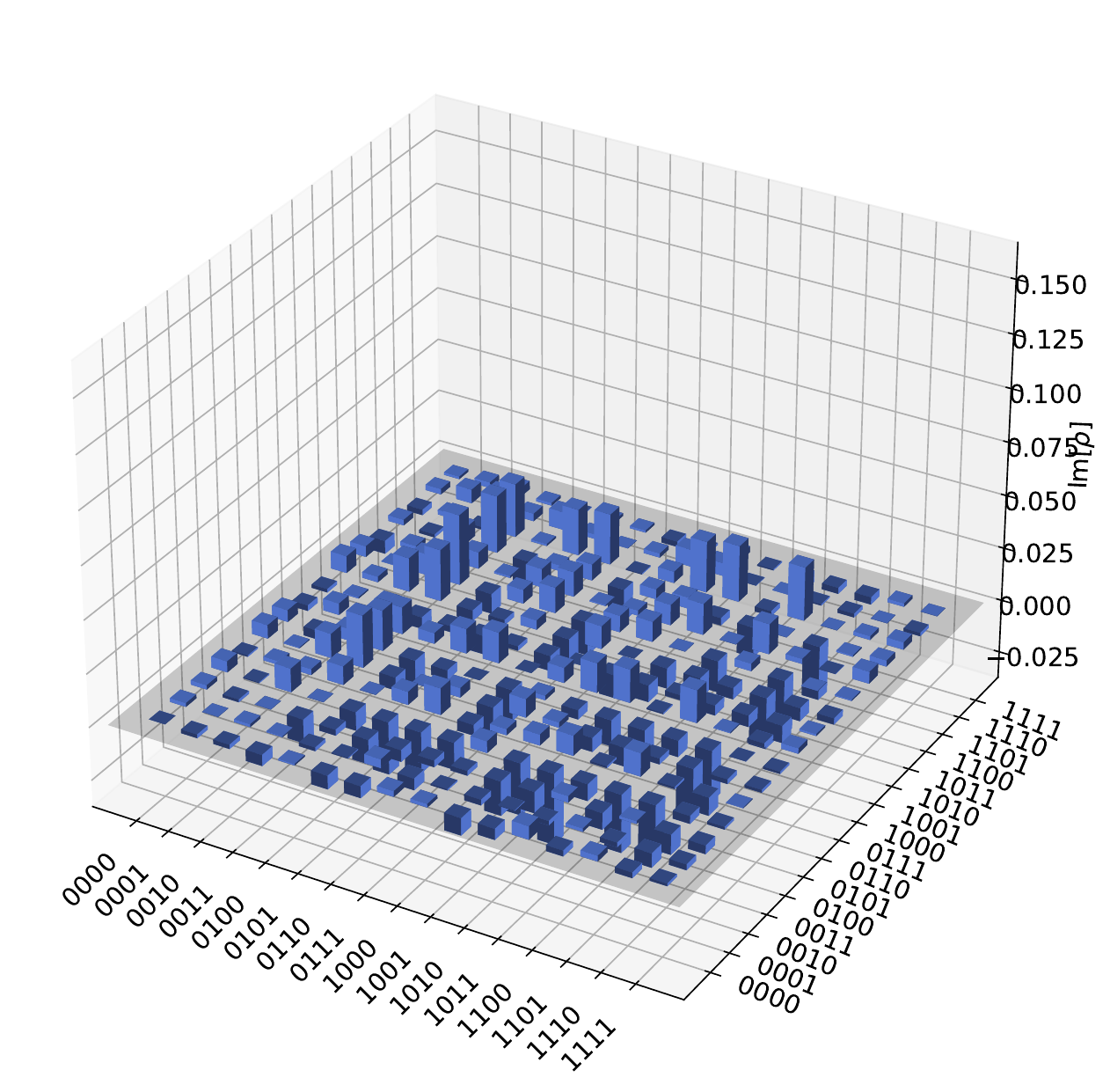}%
    \caption{State city plots of (LEFT) pure Dicke State $\dickeDM{4}{2}$, (MIDDLE) measured state $\rho$, (RIGHT) error-mitigated $\rho$.
        In each column, the (TOP) plot shows the real value of the corresponding density matrix entry $\rho_{ij}$ and the (BOTTOM) plot the imaginary part.
        The measured states were prepared on IBM~Q Montreal using state tomography circuits with an initial layout provided transpilation, 
        with and without the use of measurement error mitigation.
    }
    \label{fig:state-city-plots}
\end{figure*}

\label{sec:setup}
\subsection{Definitions}

Informally, our goal is to measure how well our circuits prepare the Dicke states $\dicke{n}{k}$ for $n\leq 6,\ k \leq n/2$ on IBM~Q devices Sydney and Montreal. To make this precise, we first introduce some notation. 
On one hand, on an ideal quantum device, our circuits would prepare the pure quantum state $\delta := \dickeDM{n}{k}$. When measured in the computational basis, the probability $q_i$ to get basis state $\ket{i}$ is
\begin{align}
    q_i =  \tr(\ketbra{i}\delta) = \langle i \dickeDM{n}{k} i \rangle   
        =  \begin{cases}
                \tfrac{1}{\binom{n}{k}} & \text{if $\mathit{\rm wt}(i)=k$,}\\
                0                       & \text{otherwise.} 
            \end{cases}
    \label{eq:ideal-prob}
\end{align}

On the other hand, due to noise on current NISQ devices, the circuits prepare a noisy mixed state~\cite{nielsen2002quantum}, 
denoted with the density matrix $\rho$, a positive semi-definite Hermitian matrix.
A computational basis measurement gives basis state $\ket{i}$ with probability $p_i := \tr(\ketbra{i}\rho) = \bra{i}\rho\ket{i} = \rho_{ii}$.
As we can neither directly observe the mixed state $\rho$ nor the probabilities $(p_i)_i$, we fit the results of quantum state tomography circuits with a maximum-likelihood estimation~\cite{james2001tomography}, 
while we use relative frequencies of sampling $\ket{i}$ to estimate the probability $p_i$. 
All of this is natively supported in QISKIT~\cite{qiskit-tomography}.

As an example, we show density matrices $\delta = \dickeDM{4}{2}$ and $\rho$ as state city plots in Figure~\ref{fig:state-city-plots}, 
with two versions of the experimentally derived $\rho$: one that we get from the state tomography circuit results, and one which we get from the results combined with measurement error mitigation, 
discussed in Subsection~\ref{subsec:experimental-setup} and in more detail in~\ref{subsec:mitigation}.

\subsubsection{Quantum Fidelity}
Quantum state fidelity is a quantitative measure on the closeness of two quantum states. In our context, it measures how close the output state $\rho$ of a Dicke state preparation circuit is to the ideal expected state $\dicke{n}{k}$. Thus the fidelity signifies the probability that $\rho$ passes the yes/no test of being the pure $\dicke{n}{k}$ (with the test being the measurement of the observable $\delta = \dickeDM{n}{k}$)~\cite{jozsa1994fidelity}.

More generally, the fidelity between $\delta$ and $\rho$ can range from 0 to 1 and is defined as%
\footnote{We note that some authors~\cite{nielsen2002quantum} use the non-squared trace as an alternative definition for fidelity,
$F'(\delta,\rho) = \tr\surd(\sqrt{\delta}\rho\sqrt{\delta})$.}  
\begin{align}
    F(\delta,\rho) = \left[\tr\left(\sqrt{\sqrt{\delta}\rho\sqrt{\delta}}\right)\right]^2,
    \label{eq:fidelity-mixed}
\end{align}
where $\sqrt{\rho}$ denotes the unique positive semi-definite square root of $\rho$ such that $\sqrt{\rho}\sqrt{\rho}=\sqrt{\rho}^{\dagger}\sqrt{\rho} = \rho$. Since $\delta$ in our case is pure and thus $\sqrt{\delta}=\delta$, Equation~\ref{eq:fidelity-mixed} reduces to 
\begin{align}
    F(\dicke{n}{k},\rho) = \dickeT{n}{k}\rho\dicke{n}{k}.
    \label{eq:fidelity-test}
\end{align}
Finally, if $\rho$ were a pure state as well, $\rho=\ketbra{\psi}$, the fidelity would reduce to the squared state overlap 
\begin{align}
    F(\dicke{n}{k},\psi) =
    \left|\dickeT{n}{k}\psi\rangle \right|^2.
    \label{eq:fidelity-overlap}
\end{align}
To determine the density matrix $\rho$ describing the mixed state prepared by our Dicke state circuits on IBM~Q devices, we run full state tomography with $3^n$ different circuits, each sampling the prepared state in a particular Pauli basis.

\subsubsection{Hellinger Fidelity}

In the case of classical probabilities (where density matrices reduce to diagonal matrices with $\delta_{ii} = q_i,\ \rho_{ii} = p_i$), 
the quantum fidelity~\eqref{eq:fidelity-mixed} reduces to an analogous measure for the similarity between two probability distributions $p=(p_i)_i$, $q=(q_i)_i$
called Hellinger fidelity
\begin{align}
    H(q,p) = \left[ \sum_{i=1}^{n}\sqrt{p_i \cdot q_i} \right]^{2}.
    \label{eq:fidelity-hellinger}
\end{align}
The Hellinger fidelity is related to maybe better known measures Hellinger distance HD $=\tfrac{1}{\surd{2}}\surd{\sum (\sqrt{p_i}-\sqrt{q_i})^2}$~\cite{hellinger1909}
and Bhattacharya coefficient BC $= \sum \sqrt{p_i\cdot q_i}$~\cite{bhattacharyya1946coefficient} through HD $=$ BC\textsuperscript{2} $=$ (1-HD\textsuperscript{2})\textsuperscript{2}.

The Hellinger fidelity between the ideal Dicke state probability distribution $q_i$ from Equation~\ref{eq:ideal-prob} and the measurement probabilites $p_i=\rho_{ii}$ for Dicke states prepared on IBM~Q only depends on the diagonal elements of the state's density matrix $\rho$, 
thus ignoring information on coherence between different basis states.
On the plus side, this circumvents the high number of tomography circuits, 
as $p_i$ can be estimated from the relative frequency of samples $\ket{i}$ 
we get from measuring the Dicke state preparation circuit in the computational basis. 

\subsubsection{Measured Success Probability}

An even more simplistic measure has been used in the literature~\cite{mukherjee2020preparing}, which is to simply estimate the probability that a measurement in the computational basis yields a state $\ket{i}$ with Hamming weight $\mathit{\rm wt}(i) = k$.
We we call this the measurement success probability:
\begin{align}
    M_k(p) = \sum\nolimits_{\mathit{\rm wt}(i)=k} p_i.
    \label{eq:fidelity-msp}
\end{align}
The measurement success probability not only drops the evaluation of coherence, but also does not discriminate between a uniform distribution
over Hamming weight $k$ states and skewed distributions with the same 
success probability.
We now explore the relation between the three measures from Equations~\eqref{eq:fidelity-test},~\eqref{eq:fidelity-hellinger} and~\eqref{eq:fidelity-msp} and existing benchmarks.
  
\subsection{Relationship of measures, benchmarks}

As we have seen, the measures quantum fidelity $F$, Hellinger fidelity $H$ and measurement success probability $M_k$ intuitively are increasingly less strict.
This is also true mathematically, i.e. we show the relation $F\leq H\leq M_k$:
\begin{align*}
    F(\dicke{n}{k},\rho) 
    &= \dickeT{n}{k}\rho\dicke{n}{k} = \dickeT{n}{k}\Id \sqrt{\rho}^\dagger \sqrt{\rho} \Id \dicke{n}{k}    \\
    &= \sum_{i=1}^n \sum_{j=1}^n \langle D_k^n \ketbra{i} \sqrt{\rho}^\dagger \sqrt{\rho} \ketbra{j} \rangle D_k^n    \\
    & \stackrel{CS}{\leq} \sum_{i=1}^n \sum_{j=1}^n \underbrace{\dickeT{n}{k}i\rangle}_{\sqrt{q_i}} \underbrace{\sqrt{\bra{i}\rho\ket{i}}}_{\sqrt{p_i}} \underbrace{\sqrt{\bra{j}\rho\ket{j}}}_{\sqrt{p_j}} \underbrace{\langle j \dicke{n}{k}}_{\sqrt{q_j}}   \\
    & = \left(\sum_{i=1}^n \sqrt{q_i p_i} \right)^{\mathclap{2}} = H(p,q) \stackrel{\eqref{eq:ideal-prob}}{=} \left(\sum_{\mathclap{\qquad\mathit{\rm wt}(i)=k}} \sqrt{q_i}\sqrt{p_i}\right)^{\mathclap{2}} \\
    & \stackrel{CS}{\leq} \left(\sum_{\mathit{\rm wt}(i)=k}q_i\right) \left(\sum_{\mathit{\rm wt}(i)=k}p_i\right) \stackrel{\eqref{eq:ideal-prob}}{=} M_k(p),
\end{align*}
where we used the Cauchy-Schwarz inequality (CS) and the $q_i$ definition~\eqref{eq:ideal-prob} twice each.\hfill$\qed$

We will be interested in how well the classical measures Hellinger fidelty and measurement success probability \emph{upper bound} the classical fidelity. 
Note that because $\rho$ and $p_i$ are only estimated through a maximum-likelihood fitter and relative frequencies of samples, respectively,
there is a small chance for statistical fluctuations violating the inequalities.

In lieu of lower bounds, we turn instead to existing benchmarks for quantum fidelities with Dicke states (see Figure~\ref{fig:fidelity-comparison}):
\begin{itemize}
    \item   quantum fidelities ranging from $\sim 0.95$ to $\sim 0.75$~\cite{epfl2019wstate} for W state preparation of $\dicke{2}{1}$, $\dicke{3}{1}$, $\dicke{4}{1}$, $\dicke{5}{1}$,
    \item   quantum fidelity of 0.53~\cite{mukherjee2020preparing} for the $\dicke{4}{2}$ preparation,
    \item   squared state overlaps~\eqref{eq:fidelity-overlap} ranging from $0.5$ to $0.3125$ for product states $(\sqrt{1-k/n}\ket{0}+\sqrt{k/n}\ket{1})^{\otimes n}$.
\end{itemize}
The latter is actually the state with highest quantum fidelity with $\dicke{n}{k}$ among all $n$-qubit product states. We will prove this formally, as it is not clear a priori: based on the symmetry of Dicke states, we know that any symmetric permutation of qubits in a product state can not change its squared state overlap with the Dicke state, however this does not imply directly that the best possible product state itself has to be symmetric. 
Let $\ket{\psi} = \bigotimes_{i=1}^n (a_i\ket{0}+b_i\ket{1})$ be any $n$-qubit product state, where $|a_i|^2+|b_i|^2=1$. 
We are interested in all terms of the state vector which have Hamming weight $k$, i.e. whose amplitudes consist of $n-k$~$a$s and~$k$~$b$s. 
Then its quantum fidelity with $\dicke{n}{k}$ satisfies
\begin{align*}
    \left|\dickeT{n}{k} \psi\rangle\right|^2
    & = \left| \frac{1}{\surd\binom{n}{k}} a_{i_1}\cdot\ldots\cdot a_{i_n} \sum_{1\leq i_1 \leq \ldots \leq i_k \leq n} \frac{b_{i_1}}{a_{i_1}} \cdots \frac{b_{i_k}}{a_{i_k}} \right|^2    \\
    & \stackrel{\mathclap{CS}}{\leq} \frac{1}{\binom{n}{k}} \left(\sum_{\mathclap{\qquad\qquad\quad1\leq i_1 \leq \ldots \leq i_{n-k} \leq n}} |a_{i_1}|^2 \cdots |a_{i_{n-k}}|^2 \right) \left(\sum_{\mathclap{\qquad\qquad1\leq i_1 \leq \ldots \leq i_k \leq n}} |b_{i_1}|^2 \cdots|b_{i_{k}}|^2 \right) \\
    & = \frac{\binom{n}{k}^2}{\binom{n}{k}} \left(\frac{\sum |a_{i_1}|^2 \cdots |a_{i_{n-k}}|^2}{\binom{n}{n-k}}\right) \left(\frac{\sum |b_{i_1}|^2 \cdots|b_{i_{k}}|^2}{\binom{n}{k}} \right) \\
    & \stackrel{\mathclap{M}}{\leq} \binom{n}{k} \left(\frac{\sum\nolimits_{i=1}^n |a_i|^2}{n}\right)^{n-k} \left(\frac{\sum\nolimits_{i=1}^n |b_i|^2}{n}\right)^{k}  \\
    & = \tfrac{\binom{n}{k}}{n^n k^{n-k} (n-k)^k} \left( k \sum|a_i|^2\right)^{n-k} \left( (n-k) \sum|b_i|^2\right)^{k}    \\
    & \stackrel{\mathclap{GA}}{\leq} \tfrac{\binom{n}{k}}{n^n k^{n-k} (n-k)^k} \left(\frac{(n-k) k\sum |a_i|^2 + k(n-k)\sum |b_i|^2}{n}\right)^{\mathclap{n}}    \\
    & = \frac{\binom{n}{k} k^n (n-k)^n}{n^n k^{n-k}(n-k)^k}
    = \binom{n}{k} \left(\frac{k}{n}\right)^k \left(\frac{n-k}{n}\right)^{n-k},
\end{align*}
using the Cauchy-Schwarz inequality~(CS), the Maclaurin inequality for elementary symmetric polynomials~(M) and the Geometric \& Arithmetic Mean inequality~(GA). In the inequalities, equality holds for Maclaurin if all $|a_i|^2$ are equal (and all $|b_i|^2$ are equal, respectively), for the Geometric \& Arithmetic Mean if $(n-k)\sum|b_i|^2 = k\sum |a_i|^2$ and for Cauchy-Schwarz if the fractions $\tfrac{b_{i_1}\cdots b_{i_k}}{a_{i_1}\cdots a_{i_k}}$ have the same value for all index choices $i_1,\ldots,i_k$. Equality along the whole chain thus holds if for all $i$ we have $a_i = \sqrt{1-k/n}$ and $b_i = \sqrt{k/n}$.\hfill \qed

\subsection{Experimental Setup}
\label{subsec:experimental-setup}
    \subsubsection{Measurement Error Mitigation: }
        Noise in gate-level quantum computers generates measurement results that are far away from the results expected from an ideal quantum computer. Unusual behavior of physical quantum gates and relaxation of quantum energy over time are considered the primary source of noise which represents a massive challenge in quantum computing. There is another form of noise that occurs during the final measurement step that affects the perfect output state and outputs randomly perturbed noisy state. It is possible to understand the effect of those measurement errors by first generating circuits for all possible $2^n$ basis states and measuring them on a real quantum system to find the probability of each states.
        Theoretically these perfect basis states becomes noisy before returning to the user as output i.e. $Output_{actual} = M_{noise} * Output_{ideal}$. If we can compute the matrix that when multiplied with the ideal state generates the measured state, it is possible to get the mitigated output. From linear algebra, the inverse of $M_{noise}$ can be applied to mitigate the measurement errors from the $Output_{actual}$ state i.e. $Output_{mitigated} = M_{noise}^{-1} * Output_{actual}$~\cite{qiskit-mitigation}.
    
    \subsubsection{Transpilation}
        We construct circuits for Dicke States $\ket{D^n_k}$ where $n = 2,3,4,5,6$ and $k = 1,2,3$ using the divide and conquer approach described in Section~\ref{sec:circuits}.  IBM QISKIT compiler (version: 0.26.2) is used to compile the logical circuits onto the specified hardware using the native gate set. As pre and post compiled circuits remain in the same language and both are equivalent in terms of unitary representation, the Qiskit compiler is termed a `transpiler' while the compilation is termed as `transpilation'~\cite{qiskit-transpiler}. QISKIT offers different transpilation \& optimization options that generate custom transpiled circuits. For this experiment, we set the QISKIT optimization level to 3 and use three different transpilation options:
            i) \textbf{Default}: We only use the default options. 
            ii) \textbf{Initial Layout Provided}: We provide a subgraph of the required number of qubits as the initial layout where we choose the qubits focusing on LNN connectivity \& gate error rate. 
        iii) \textbf{Noise Adaptive}: We set \textit{layout\_method} argument as  \textit{noise\_adaptive} and depend on QISKIT to select qubits. In the noise adaptive mapping, the qubits are selected using the specific machine topology and daily calibration data from IBM. It avoids qubits with high error rates and low coherence times to enable maximum reliability for {\CNOT}s and reduce qubit movements \cite{murali2019noise}.

    \subsubsection{Evaluation}
        To verify the circuits, we compute the quantum fidelity to measure the closeness of the prepared state $\rho$ to the Dicke state $\dicke{n}{k}$. To calculate $\rho$, we first generate state tomography circuits in each measurement combination of the Pauli X, Y, and Z bases for $n$-qubits that give a total of $3^n$ measurement circuits.  We run the tomography circuits in IBM Q systems and calculate fidelity between pure output state and tomography circuit measurement. For this experiment, we selected IBM Q Sydney and IBM Q Montreal quantum systems from the IBM Quantum Experience cloud service. IBM Q Sydney and IBM Q Montreal have the same elongated hexagon topology on their 27 qubits. The quantum volume of IBM Q Sydney is 32 and the quantum volume of IBM Q Montreal is 128. The average {\CNOT} gate error in IBM Q Sydney \& IBM Q Montreal is 0.0114 and 0.1522 where the average readout error is 0.05078 and 0.02028, respectively.
        
        We apply measurement calibration to mitigate noise in the final measurement steps of the tomography circuits.  We use the QISKIT ignis package to mitigate the noise by first creating a measurement filter object, then applying the filter to the actual output state. Then we compute fidelity again between the ideal state and error mitigated output state to compare with the previous fidelity. 
        
        We also  compare quantum fidelity with Hellinger fidelity and measured success probability for different transpilation options. For each experiment, we run all $3^n$ tomography circuits using `$qiskit.execute$' where all the cirucits are executed together, alternatively also using `$IBMQJobManager.run$' where we split the circuits into multiple jobs and set to execute a fixed number of circuits in a job.
        
\section{Results}
\label{sec:results}
   We explain experimental results on IBM Q Sydney and Montreal systems for the different Dicke States using the transpilation options described in Subsection~\ref{subsec:experimental-setup}. We denote the Dicke State circuits as $Dnk \mhyphen r$ where $n = \{2,3,4,5,6\} $ represents the number of qubits, $k = \{2,3,4,5,6\}$ represents the Hamming weight, and $r$ denotes the number of {\CNOT} gates in the circuit. 
   All in all, the Dicke states $\dicke{1}{2},\dicke{1}{3},\dicke{1}{4},\dicke{1}{5},\dicke{1}{6},\dicke{2}{4},\dicke{2}{5},\dicke{2}{6}$ and $\dicke{3}{6}$ are prepared 
   by circuits $\dickep{2}{1}{1}$, $\dickep{3}{1}{2}$, $\dickep{4}{1}{5}$, $\dickep{5}{1}{7}$, $\dickep{6}{1}{9}$, $\dickep{4}{2}{10}$, $\dickep{5}{2}{17}$, 
   $\dickep{6}{2}{24}$, $\dickep{6}{3}{32}$, $\dickep{6}{3}{33}$. The experiments using `$qiskit.execute$' are denoted as `$All \; circuit$' and experiments with
   `$IBMQJobManager.run$' are denoted as `$Split\;jobs$'. 

       \begin{figure*}[t!]
            \centering
            \includegraphics[width=0.99\linewidth,height=.72\linewidth] {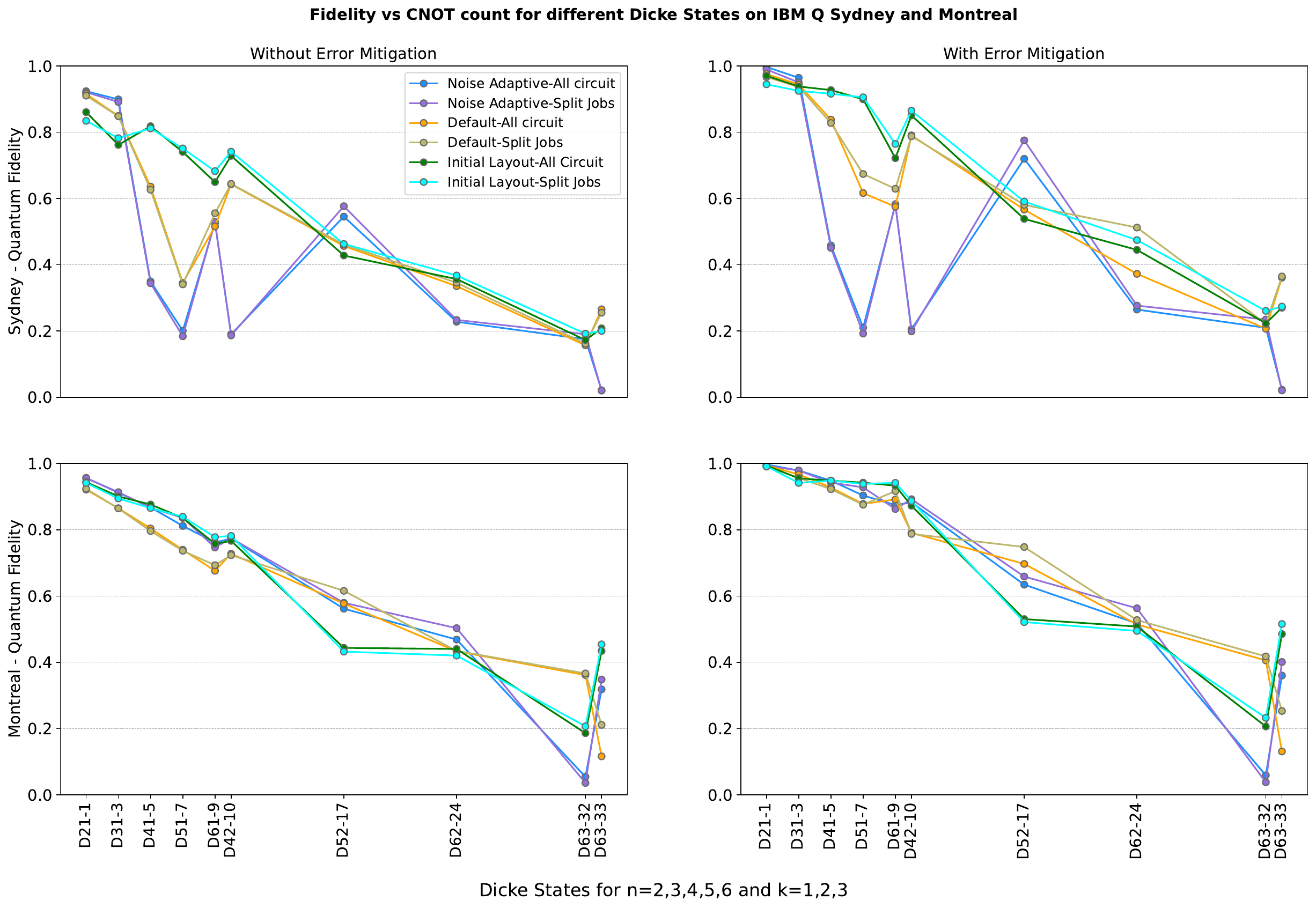}
            \caption{
            { Measured fidelity of Dicke States for IBM Q Sydney system without error mitigation (top left) \& with error mitigation (top right) and IBM Q Montreal system without error mitigation (bottom left) \& with error mitigation (bottom right) using different transpilation method of QISKIT. We varied default, noise adaptive and initial layout provided transpilation of QISKIT with optimization level 3. Additionally we applied error mitigation techniques of QISKIT to reduce measurement noise from the output.  }}
            \label{Fig:fidelity-vs-cnot}
    \end{figure*}  
        
    \subsection{Quantum State Fidelities}

        We first study quantum state fidelities with respect to the number of {\CNOT} gates required for all Dicke State circuits $Dnk \mhyphen r$ on different quantum backends. Figure \ref{Fig:fidelity-vs-cnot} shows measured quantum fidelity of all Dicke States on IBM Q Sydney and IBM Q Montreal system for different transpilation options of IBM QISKIT. Each plot has six lines representing different transpilation techniques and job execution methods outlined in Section \ref{sec:setup}. The Dicke States are sorted (left to right) according to the number of required {\CNOT} gates in the \emph{untranspiled} circuit. The plots show that for almost all experiments, `$Split  \;jobs$' is equal to or slightly better than `$All\;circuit$' execution; for these experiments, `$Split\;jobs$' option has fewer circuits than `$All\;circuit$' per job.

        We present how fidelity changes if we apply measurement error mitigation techniques to the measurement result for each experiment. The left two plots show quantum fidelities on Sydney \& Montreal without error mitigation techniques and the right two plots show quantum fidelities after applying error mitigation techniques. We observe that for both Sydney \& Montreal,  measurement error mitigation reduces noise from the output and increases quantum state fidelities irrespective of circuit length and width.
        
        Theoretically, the number of {\CNOT} count and quantum fidelity measure should have a negative correlation i.e. when {\CNOT} count increases, fidelity measure decreases, and vice versa. We observe that fidelities for Montreal follow this trend for all states with chosen transpilation methods except for state D63-32 where the fidelities are less than state D63-33. From our circuit design, the circuit depth of state D63-33 is (depth = 29) is less than the circuit depth of state D63-32 (depth = 33). So, we can say that reduction in circuit depth increases parallel execution of gates and thus decreases the overall error rate. However, we see from the plot that circuits using default transpilation do not obey this trend for the Montreal backend. We also note that fidelities for Sydney are more distorted especially for noise adaptive transpilation of circuits D41-5, D51-7, D61-9, and D42-10. We investigate the circuit after noise adaptive transpilation and find that the chosen qubits are sometimes not connected which adds more ancilla qubits and thus increases the number of gates. Besides, we find almost similar characteristics for IBM Q Sydney and Montreal using the initial layout provided and default transpiled circuits.  
        
        Overall, IBM Q Montreal behaves better with better fidelities than Sydney and there is no irritating ups and down like Sydney. Thus it can be said that the newer system (Montreal) is better and it indeed deserves better Quantum Volume (QV) ranking. Fidelity values of more than 0.5 give us high confidence that the IBMQ devices indeed manage to create entanglement in the Dicke states.

    \subsection{Influence of Measurement Error Mitigation}
    \label{subsec:mitigation}
    
           \begin{figure*}[t!]
            \centering
            \includegraphics[width=0.99\linewidth] {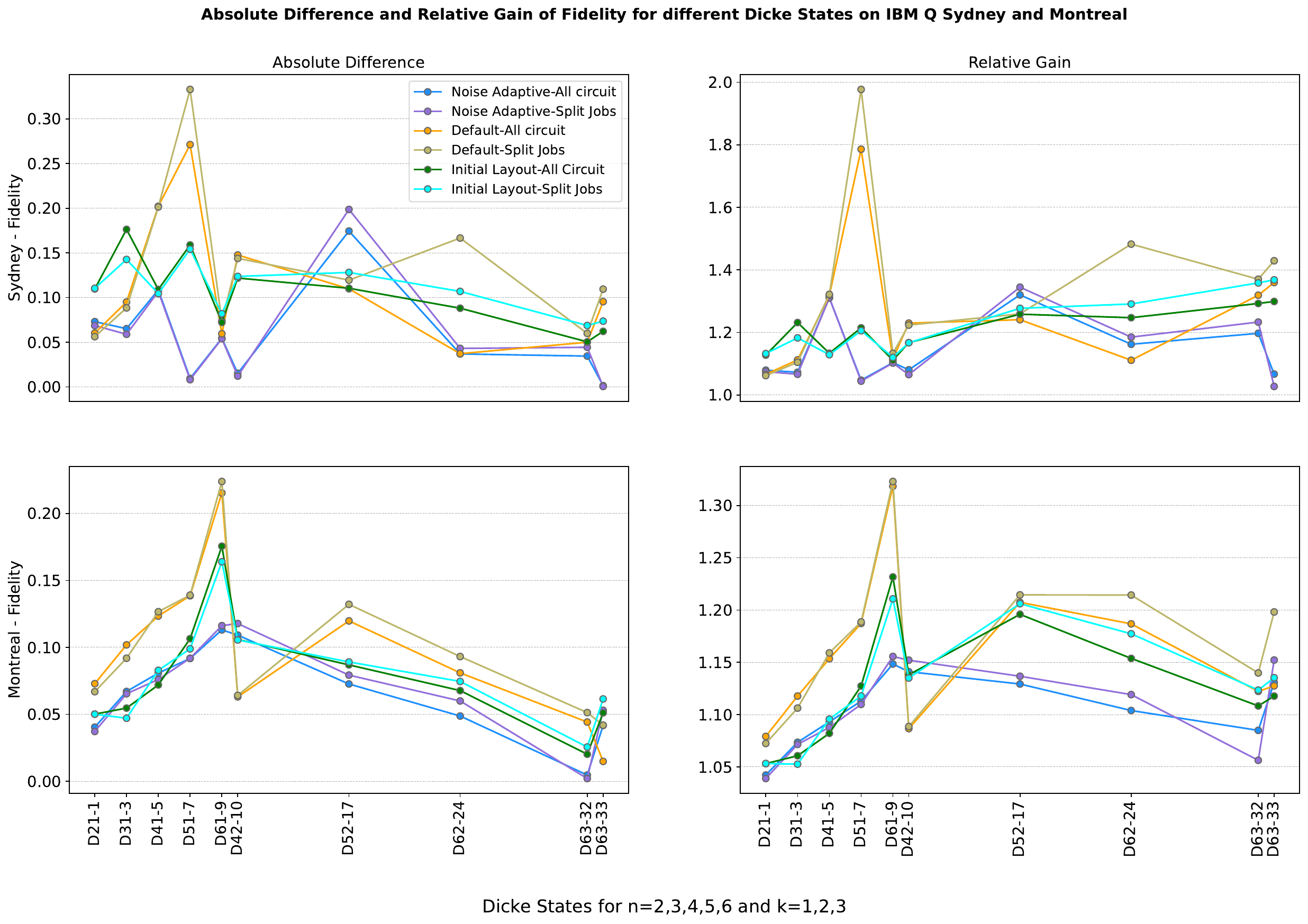}
            \caption{
            { Absolute difference and relative gain of quantum fidelity Dicke States fidelities on IBM Q Sydney (left) and IBM Q Montreal system (right) are measured using different transpilation method of QISKIT with and without applying error mitigation techniques.}}
             \label{Fig:absdiff-gain-fidelity}
        \end{figure*}    
        
        To investigate the effect of measurement error mitigation on output, first, we present the absolute difference of quantum fidelities before and after applying error mitigation techniques on IBM Q Sydney \& Montreal backend in Figure \ref{Fig:absdiff-gain-fidelity}. Similarly, Figure \ref{Fig:absdiff-gain-fidelity} shows the gain of quantum fidelities before and after applying error mitigation techniques on those backends. Six lines are representing different transpilation options and execution methods. All Dicke states denoted using the form Dnk$\mhyphen$r are sorted according to number of {\CNOT} gates of untranspiled circuit. 
        
        From Figure \ref{Fig:absdiff-gain-fidelity}, we observe that in most of the cases, with the increasing number of the qubits, the gain of quantum fidelity increases. Additionally, for the same value of $n$, fidelity gain is higher for smaller $k$. This is expected because, for the same $n$, the higher value of $k$ represents a more complex circuit that has more circuit noise than measurement noise. So, when we apply measurement error mitigation on a less complex circuit (smaller $k$) for the same $n$, we get a higher fidelity gain. In general, circuits with an increasing number of qubits get more benefited from measurement error mitigation. For states with the same circuit length, circuit noise dominates over measurement noise in a more complex circuit. However, we observe some exceptions in IBM Q Sydney \& Montreal for noise adaptive and initial layout provided transpilation option.

    \subsection{Info from Classical Distribution}
        Theoretically, quantum fidelity is upper bounded by the Hellinger fidelity, which in turn is upper bounded  by the success probability of output states i.e $Quantum\;Fidelity \leq Hellinger\;Fidelity \leq Measured\;Success\;Probability$. As we discussed before, quantum fidelity requires $3^n$ tomography circuits which can be costly and time consuming. Another approach could be to calculate Hellinger fidelity and measured success probability to get an estimation of quantum fidelity of our circuit. In this section, we will explore a comparative study of measured success probability, Hellinger fidelity, and quantum fidelity for our proposed Dicke state preparation circuits. From this section, we only present results from our `$All\;Circuit$' execution to avoid complexity.

\begin{figure*}[t]
            \centering
            \includegraphics[width=0.99\linewidth] {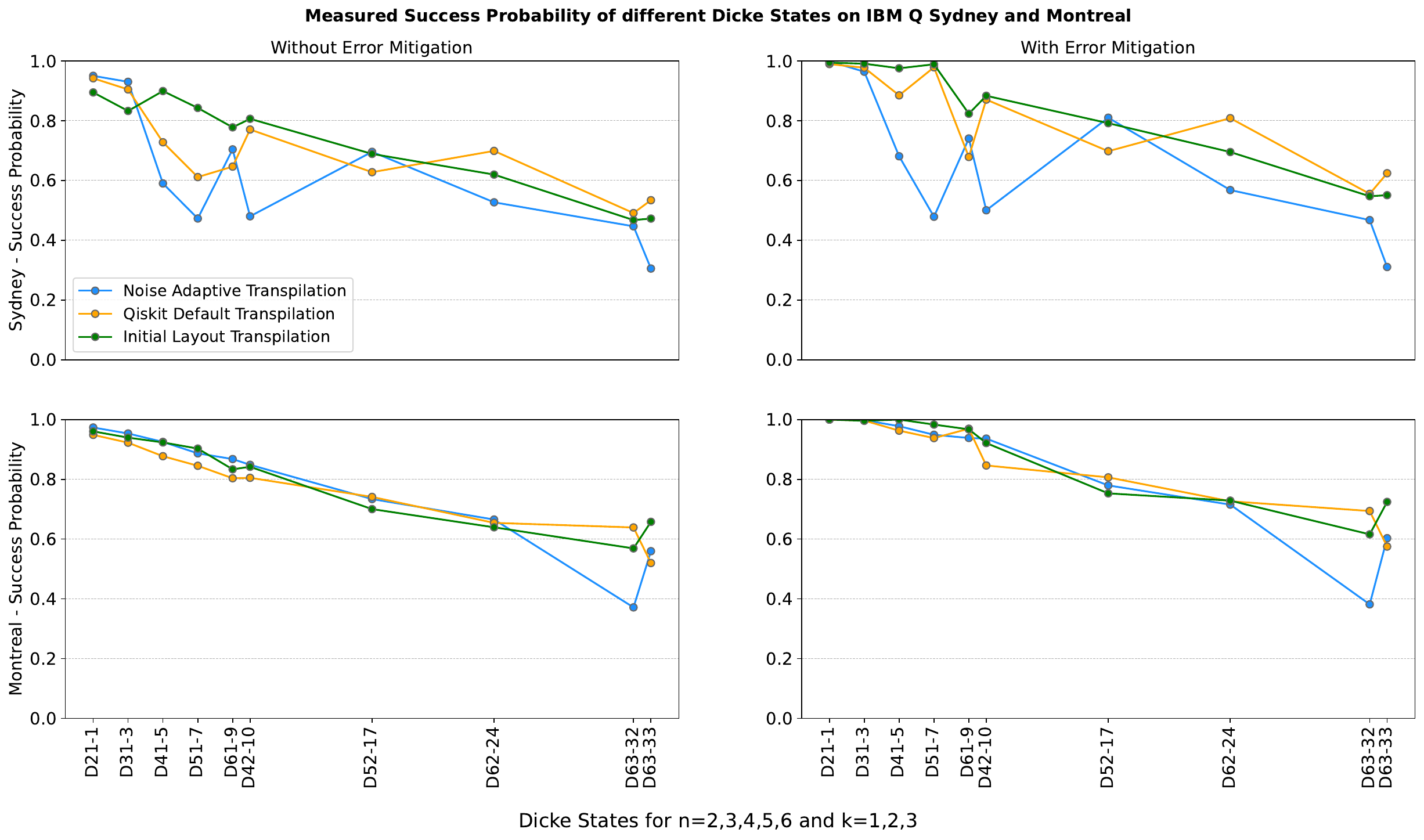}
            \caption{
	    { Measured Success Probability  of different Dicke States on IBM Q Sydney without error mitigation (top left) \& with error mitigation (top right) and on IBM Q Montreal  without error mitigation (bottom left) \& with error mitigation (bottom right) using QISKIT's  different transpilation with optimization level 3.}}
            \label{Fig:sucess-prob}
        \end{figure*}

\begin{figure*}[t]
            \centering
            \includegraphics[width=0.99\linewidth] {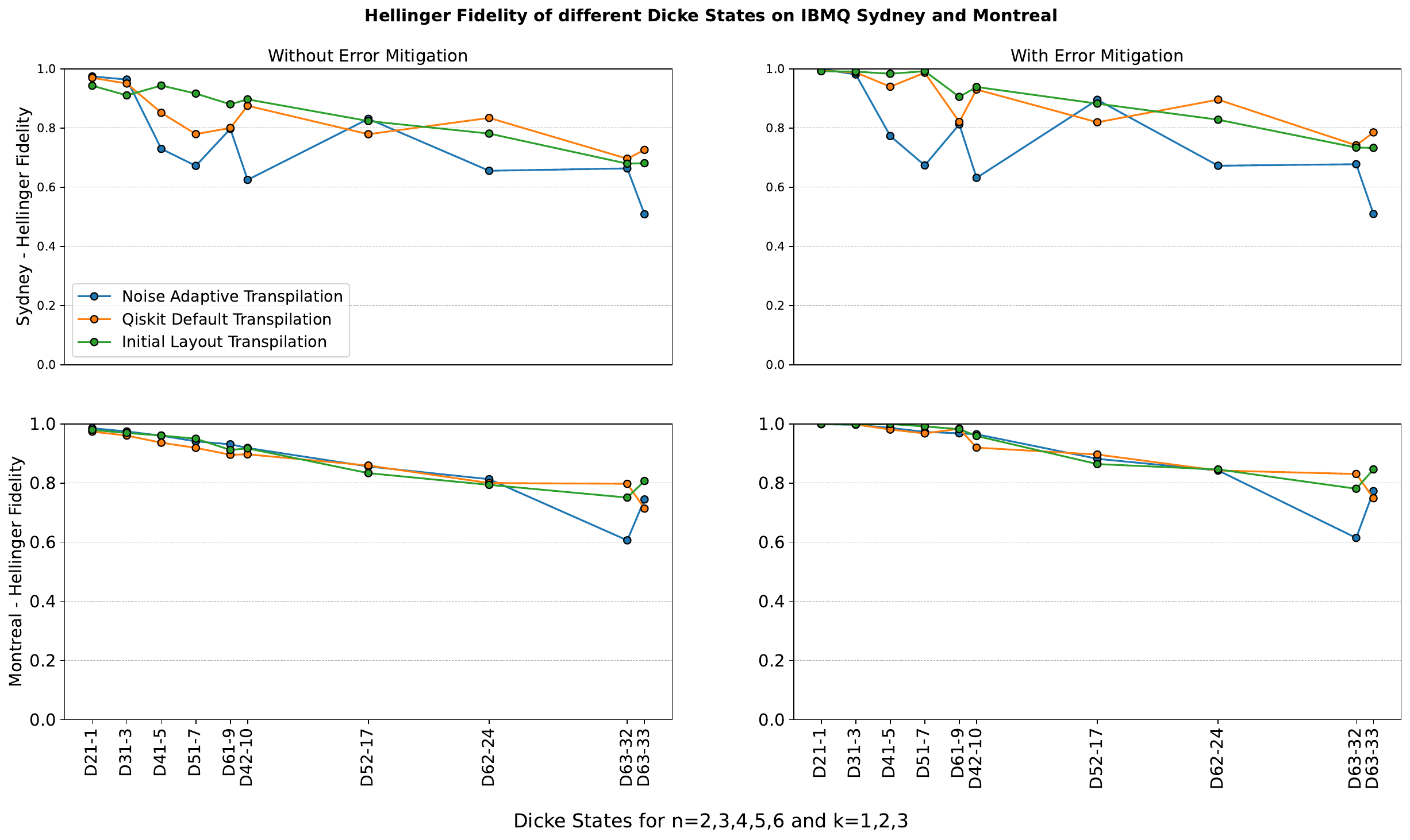}
            \caption{
            { Hellinger fidelity of Dicke States for IBM Q Sydney system without error mitigation (top left) \& with error mitigation (top right) and IBM Q Montreal system without error mitigation (bottom left) \& with error mitigation (bottom right) using different transpilation method of QISKIT.}}
             \label{Fig:hellinger-fidelity}
        \end{figure*}

        Figure \ref{Fig:sucess-prob} \& \ref{Fig:hellinger-fidelity} shows measured success probability and Hellinger fidelity of different Dicke States on IBM Q Sydney and Montreal backend. We observe that measurement error mitigation increases measured success probability and Hellinger fidelity for both backends. We can find a linear decrease of success rate and Hellinger fidelity in the plot for states sorted (left to right) according to the number of required {\CNOT} gates. However, there are some outlier points in both the plots especially for noise adaptive transpilation on the IBM Q Sydney backend. Recall that noise adaptive transpilation prioritizes error rates for selecting qubits that sometimes choose non-neighboring qubits. This induces overhead of ancilla qubits in the circuit for multi-qubits gates such as {\CNOT} which leads to the lower count of the correct output state. By definition, success probability depends on the number of correct states among a total number of shots, and Hellinger distance is calculated using the probabilities from correct and incorrect states. \\

\begin{figure*}[tbp]
            \centering
            \includegraphics[width=0.99\linewidth,height=.6\linewidth] {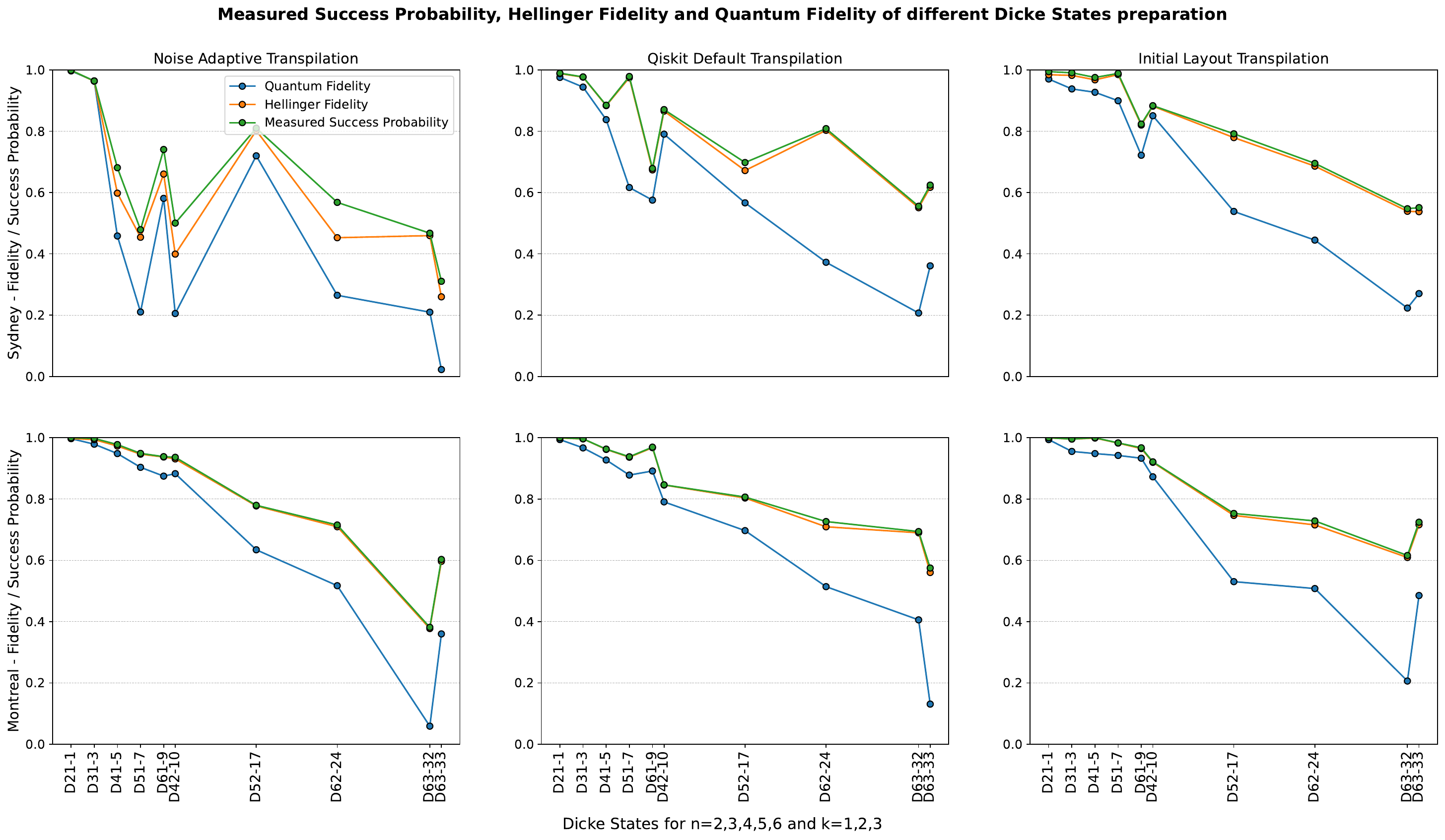}
            \caption{
            { Measured Success Probability, Hellinger Fidelity and Quantum Fidelity  of different Dicke States on IBM Q Sydney (upper three) and on IBM Q Montreal (bottom three)  with error mitigation using QISKIT's  different transpilation with optimization level~3.}}
            \label{Fig:sucess-hellinger-quantum}
        \end{figure*}

        In Figure \ref{Fig:sucess-hellinger-quantum}, we present all three measures i.e measured success probability, Hellinger fidelity and quantum fidelity for Dicke states D21-1, D31-3, D41-5, D51-7, D61-9, D42-10, D52-17, D62-24, D63-32 and D63-33 using our proposed optimized circuits. We observe that measured success probability is close to Hellinger fidelity for all transpiled circuits on IBM Q Sydney and Montreal. While quantum fidelity is close to the other two measures for the smaller circuits, with increasing circuit length and depth quantum fidelity starts to drop significantly. In summary, we find from our experiment that measured success probability and Hellinger fidelity follows theoretical expectation, and quantum fidelity increases gap with other two measure for a complex circuit. Also, the computationally inexpensive Hellinger fidelity and measured success probability are good approximations for quantum fidelity, particularly up to 4 qubits and 10 {\CNOT}s; unfortunately, the gap becomes larger for larger circuits, but the inexpensive measure still capture all the interesting trends. Note in particular, how  Hellinger fidelity replicates the behavior of quantum fidelity for the $\dicke{6}{3}$ inversions under different transpilation schemes.
        
\begin{figure*}
        \begin{center}
        \textbf{Success Probabilities of Correct and Incorrect Hamming Weight States for Dicke States \linebreak ( $n = 2,3,4,5,6$ and $k = 1,2,3)$ }\par\medskip
        \end{center}
        \textbf{\small Hamming weight, $k=1$ : } \par\medskip
            \includegraphics[width=.20\linewidth,height = .23\linewidth]{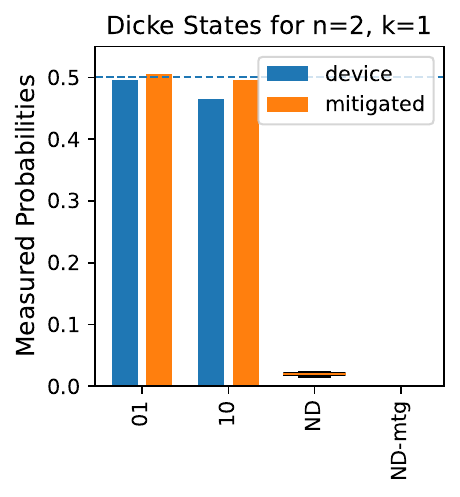}\hfill
            \includegraphics[width=.20\linewidth,height = .23\linewidth]{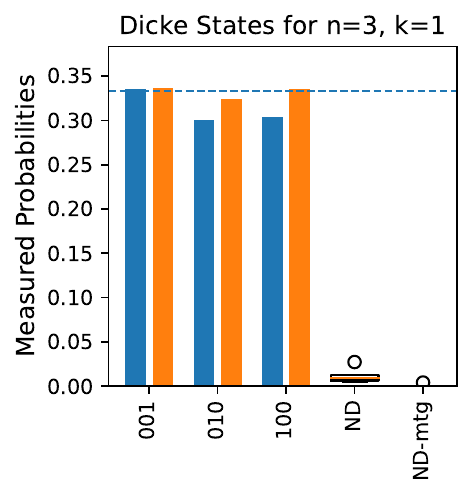}\hfill
            \includegraphics[width=.20\linewidth,height = .23\linewidth]{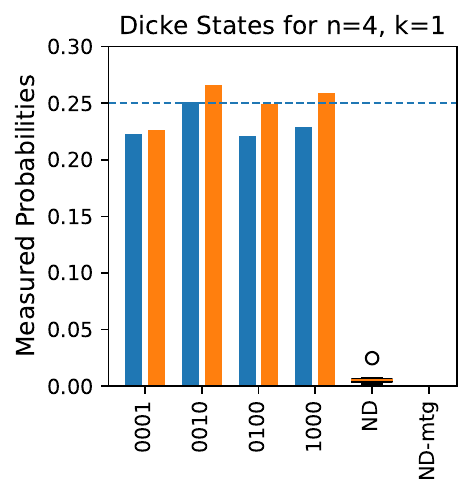}\hfill
            \includegraphics[width=.20\linewidth,height = .23\linewidth]{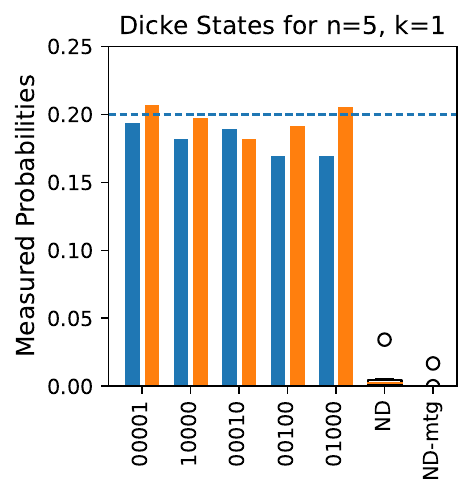}\hfill
            \includegraphics[width=.20\linewidth,height = .23\linewidth]{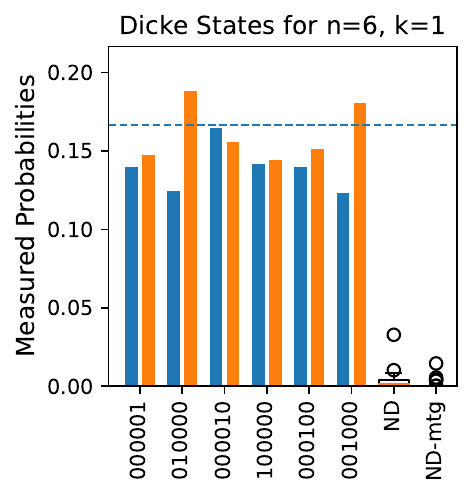}\hfill
            \\[\smallskipamount]
            
        \textbf{\small Hamming weight, $k=2$:} \par\medskip
            \includegraphics[width=.3\textwidth,height = .23\linewidth]{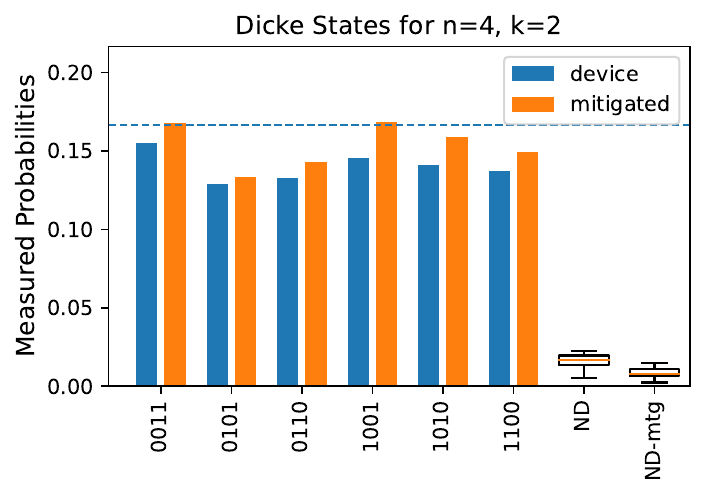}\hfill
            \includegraphics[width=.34\textwidth,height = .23\linewidth]{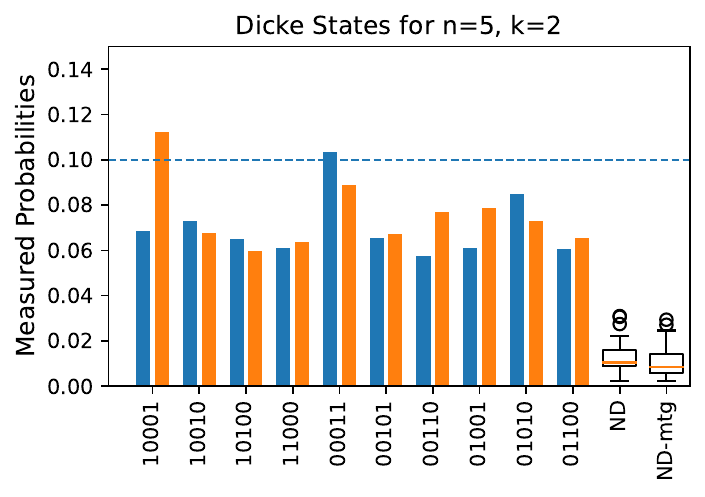}\hfill
            \includegraphics[width=.35\textwidth,height = .23\linewidth]{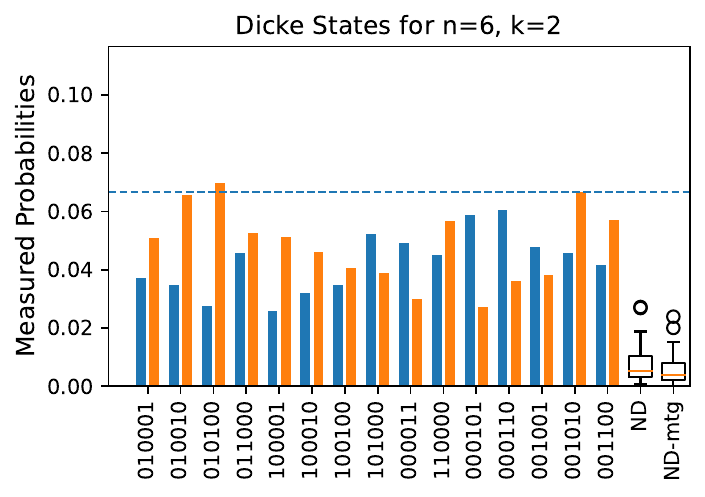}\hfill
            \\[\smallskipamount]
        \textbf{\small Hamming weight, $k=3$:} \par\medskip 
            \includegraphics[width=.5\textwidth,height = .23\linewidth]{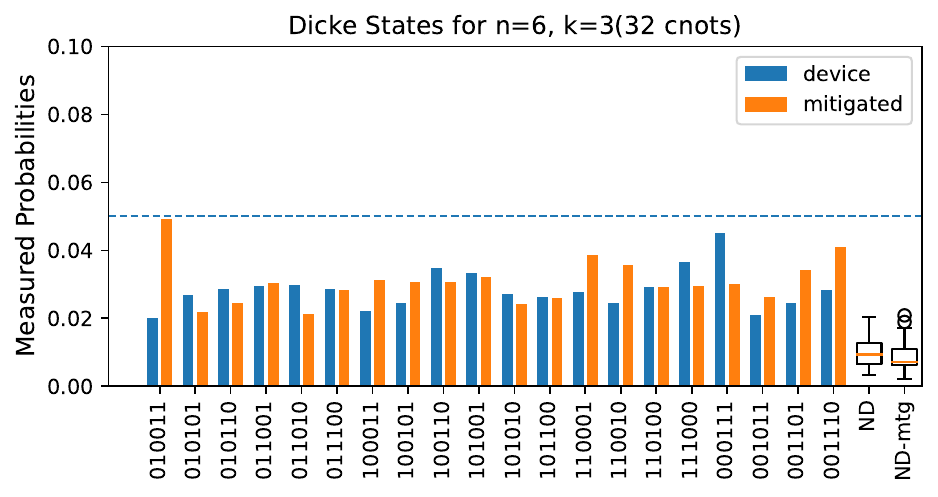}\hfill
            \includegraphics[width=.5\textwidth,height = .23\linewidth]{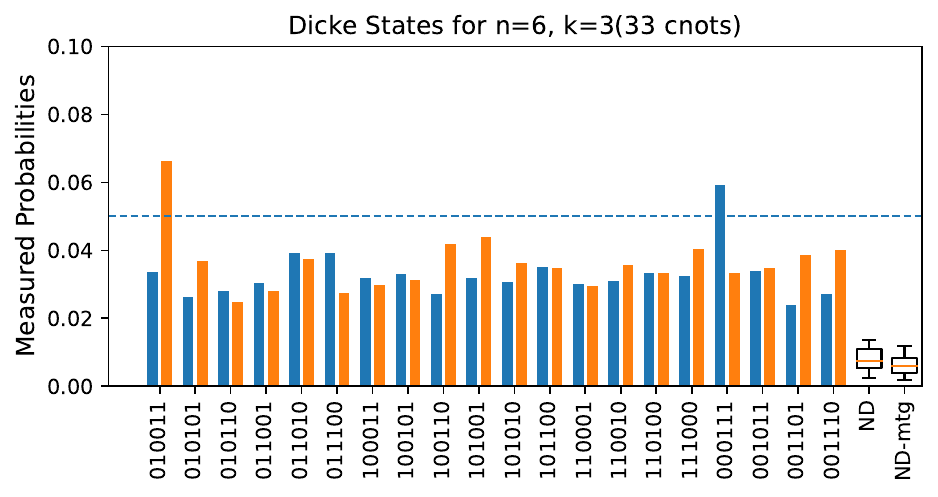}
        
            \caption{Measured Probability of correct Hamming weights vs. incorrect Hamming weight distribution for Dicke states for $n = 2, 3, 4, 5$ and $k =1, 2, 3$ using initial layout provided transpilation with optimization level 3. The results are obtained from our experiments in IBM Q Montreal device. Here, `ND' denotes the distribution of incorrect Hamming weight states (Non-Dicke). The `blue' bar represents success rate without error mitigation while `red' bar represents success rate with error mitigation. The horizontal line in each plot shows ideal probability of dicke states with correct Hamming weight. }
            \label{Fig:correct-incorrect-hw}
        \end{figure*}

        The measured success probability shown in Figure \ref{Fig:sucess-prob} represents the probability of generating the Dicke States with correct Hamming weight. It would be intriguing to observe the success probabilities of the correct and incorrect Hamming weight of a Dicke States. Figure \ref{Fig:correct-incorrect-hw} shows the measured success probability of correct Hamming weight states as well as a boxplot of incorrect Hamming weight states for our proposed Dicke States. For each plot, we show both device-generated probabilities and measurement error mitigated probabilities. The horizontal bar in each plot represents the expected probability of correct Hamming weight states i.e. $1/\binom{n}{k}$. We plot the Dicke States with the same Hamming weight ($k =1, 2, 3$)  in a row. For Hamming weight $k=1$, we see the probabilities of correct Hamming weight to be almost equal to the theoretical expectation while the incorrect Hamming weight states have a  small distribution. We find almost similar characteristics for Hamming weight $k=2$ although the measured probability decreases with increasing $n$. The success probabilities of correct Hamming weight states for Hamming weight $k=3$ are much less than the expected probabilities because circuit noise dominates over measurement noise for a complex circuit. However, the probability distribution of incorrect Hamming weight states is still very small. The plots in Figure \ref{Fig:correct-incorrect-hw} are totally in line with our results from Figure \ref{Fig:sucess-prob},  \ref{Fig:hellinger-fidelity} \& \ref{Fig:sucess-hellinger-quantum} and justifies our findings.
        
\section{Conclusion and Outlook}
    We have presented a new divide-and-conquer-style quantum algorithm for creating Dicke states, which are by definition highly entangled quantum states. Our algorithms leverage the underlying topology of the actual quantum hardware, such as linear nearest neighbor. As an additional theoretical contribution, we showed a bound of how close any product state can get to a Dicke state in terms of quantum fidelity. Our main experimental results show that present-day NISQ devices (namely two of the IBMQ backends) are indeed able to create Dicke states at very high quantum fidelity levels of more than $0.5$ at six qubits, which we verified through full state tomography, thus outperforming previously reported results. This improved performance is due to both improvements in NISQ hardware as well as our algorithmic improvements. In addition, we have examined the suitability of simpler measures of success probability and Hellinger fidelity as approximations for quantum fidelity; we found that these measures match the trends for quantum fidelity. We have also explored the IBM QISKIT software stack with respect to different compiler options and found that no single setting finds the optimum solutions in each case, thus forcing the user to experiment with different settings for their own experiments.

    We see the following future directions:
    \begin{itemize}
        \item   The cost of the ``Divide'' part scales well on Ladder architectures but not for LNN architectures. What is the trade-off of the Divide-and-Conquer approach versus architectural constraints? Can we get sub-linear circuit depths $o(n)$ for full connectivity or small constant factors for the {\CNOT} count in LNN connectivities?
        \item   The cost of full state tomography gets prohibitively large even for small $n \geq 6$. Can we approximate $\rho$ and the quantum fidelity from below with fewer tomography circuits in addition to the upper bound given by the Hellinger fidelity?
        
        \item The divide-and-conquer approach presented in this paper improved on constants in the running time~\cite{baertschi2019deterministic} and on connectivity requirements~\cite{mukherjee2020preparing}.
        To get an asymptotic improvement in the circuit depth, one may recursively apply the divide-and-conquer approach, which is active work in progress~\cite{baertschi2022short}.
        
        \item Dicke states live  in the symmetric subspace of the full Hilbert space. By considering the symmetry of Dicke states, there might be more error mitigation potential that can be developed to further improve the fidelity.
    \end{itemize}

\bibliographystyle{plainurl}
\bibliography{main}

\end{document}